\documentclass[prfluids,aps]{revtex4-2}

\usepackage{amssymb}
\usepackage{amsmath}
\usepackage{bm}

\usepackage[utf8]{inputenc}

\usepackage{graphicx}

\usepackage{xcolor}         % Colored text
\usepackage[normalem]{ulem} % Underlining
\usepackage{siunitx}        % Pretty units formatting
\usepackage{hyperref}

\newcommand{\Fig}[1]{Fig.~\ref{#1}}
\newcommand{\RRef}[1]{Ref.~\cite{#1}}
\newcommand{\Refs}[1]{Refs.~\cite{#1}}

\renewcommand{\sb}[1]{_{\text {#1}}}  %% sub-   for lower case
\def\Sb#1{_{\scriptscriptstyle\rm{#1}}}

\newcommand{\B}[1]{{\bm{#1}}}%% Bold Roman & Greek Lower & Upper Case

\begin{document}
\title{On the investigation of properties of  superfluid $^4$He turbulence using a hot-wire signal}

\author{P. Diribarne}
\author{M. Bon Mardion}
\author{A. Girard}
\author{J.-P. Moro}
\author{B. Rousset}
\affiliation{Univ. Grenoble Alpes, IRIG-DSBT, F-38000 Grenoble, France}
\author{F. Chilla}
\author{J. Salort}
\affiliation{Univ Lyon, ENS de Lyon, Univ Claude Bernard, CNRS, Laboratoire de Physique, Lyon, France}
\author{A. Braslau}
\author{F. Daviaud}
\author{B. Dubrulle}
\author{B. Gallet}
\author{I. Moukharski}
\author{E.-W. Saw}
\affiliation{Laboratoire SPHYNX, CEA/IRAMIS/SPEC, CNRS URA 2464, F-91191 Gif-sur-Yvette, France}
\author{C. Baudet}
\affiliation{CNRS, LEGI, F-38041 Grenoble, France}
\author{M. Gibert}
\author{P.-E. Roche}
\author{E. Rusaouen}
\affiliation{Univ. Grenoble Alpes, Institut NEEL, F-38042 Grenoble, France}
\author{Andrei Golov}
\affiliation{Department of Physics and Astronomy, The University of Manchester, Manchester MP13 9PL, United Kingdom}
\author{Victor L'vov}
\affiliation{Department of Chemical and Biological Physics, Weizmann Institute of Science, 7610001 Rehovot, Israel}
\author{Sergey Nazarenko}
\affiliation{Institut de Physique de Nice, Université Nice-Sophia Antipolis, Parc Valrose, 06108 Nice, France}

\begin{abstract}
We report hot-wire measurements performed in two very different, co- and
counter-rotating flows, in normal and superfluid helium at \SI{1.6}{K},
\SI{2}{K}, and \SI{2.3}{K}. As recently reported, the power spectrum of
the hot-wire signal in superfluid flows exhibits a significant bump at high
frequency (\textcite{Diribarne21}). We confirm that the bump frequency does
not depend significantly on the temperature and further extend the previous
analysis of the velocity dependence of the bump, over more than one decade
of velocity. The main result is that the bump frequency depends on the
turbulence intensity of the flow, and that using the turbulent Reynolds
number rather than the velocity as a control parameter collapses results
from both co- and counter-rotating flows.
The vortex shedding model  previously proposed, in its current form,
does not account for this observation. This suggests that the
physical origin  of the bump is related to the small scale turbulence properties
of the flow. We finally propose some qualitative physical mechanism by
which the smallest  structures of the flow, at intervortex distance,
could affect the heat flux of the hot-wire.
\end{abstract}

\maketitle

\section{Introduction}
One of the main questions in the theory of turbulence is how energy
is distributed over length scales, i.e. what is, in the $k$-space,
the energy spectrum ${\cal E} (k)$.
It is generally believed that for mechanically driven quantum
turbulence, the quantization of vortex circulation is
unimportant at scales greater than the mean distance between the superfluid
vortex lines, $\delta$, simply called intervortex distance hereafter. Thus,
the kinetic energy spectrum at such scales is distributed similarly to the
one in classical turbulence. For example, in homogeneous isotropic
fully-developed turbulence one expects  classical Kolmogorov-1941 (K41) spectrum
${\cal E} \Sb{K41}(k) \propto k^{-5/3}$ and this is actually what
measurements in turbulent superfluid flows show\cite{Maurer98,Salort10}.

In inertially driven flows, the main differences between the quantum
and classical turbulence is expected to arise at  scales smaller than $\delta$.
However, accessing both the large and the small scale parts of the
spectrum simultaneously is an experimental challenge. Large 
devices, such as SHREK~\cite{Rousset14}, help solving part of the problem
by providing a way to have both developed turbulence and still reasonably
large inter-vortex length scales, of the order of  a hundred  micron
at the smallest  Reynolds number. Still, in those conditions, the
proven Eulerian velocity  and vorticity sensors operating in He~II are
in resolution limits.

For example, in a recent paper, \textcite{Salort21} have analyzed velocity
spectra, obtained in the SHREK von K\'arm\'an
apparatus~\cite{Rousset14}, based on cantilever and ``Pitot tube'' signals.
They reported two different kinds of behavior associated to normal
and superfluid conditions, in the limit of very
low velocities, where the sensors had a sufficient temporal and
spacial resolution to resolve the high $k$ end of the Kolmogorov spectrum.
They used a hot-wire as a reference anemometer in He~I,
where its behavior is perfectly understood. Hot-wires can be designed
to have suitable temporal and spatial
resolution, (see, e.g., \Refs{Vallikivi14,Fan15,Diribarne19,LeThe21}),
but the interpretation of their signal in He~II is a
challenge~\cite{Duri15,Diribarne21}.
The main stumbling block is the apparition of a spectral bump at high frequency.
\textcite{Diribarne21} have shown that the bump in the spectral domain
is in fact the result of quasi-periodic enhanced heat flux events,
called ``glitches''. The physical origin of those glitches
is still not understood but the authors proposed two main leads: (i) the
shedding of large scale structures associated to the destabilization of the
thermal pattern that forms around the wire, and (ii) the interaction between
the thermal boundary layer and the enhanced velocity fluctuations at scales
comparable to the intervortex distance. The former is only related, at first
order, to the surrounding flow mean velocity, while the latter is expected to
depend on the turbulent properties of the flow.

In the present paper, we analyze the signal obtained from a hot-wire
in He~II and compare it to the velocity measurements performed with a
dynamic pressure anemometer (named ``Pitot tube'' hereafter)
in order to arbitrate between those leads and eventually propose alternatives
to understand the physical origin of the glitches. We take advantage
of the versatility of the SHREK apparatus to submit the hot-wire to two
main flow configurations, with very different turbulent properties.

The paper is organized as follows: after a presentation of the experimental
setup and the different flow configurations in Sec.~\ref{sec:setup}, we show
the typical shape of the spectra obtained in He~I and He~II in
Sec.~\ref{sec:spectra} and finally the velocity dependence of the spectral bump
frequency is discussed in Sec.~\ref{sec:discussion}.

\section{Experiment description}
\label{sec:setup}
\subsection{Experimental apparatus}

\begin{figure}
  \includegraphics[]{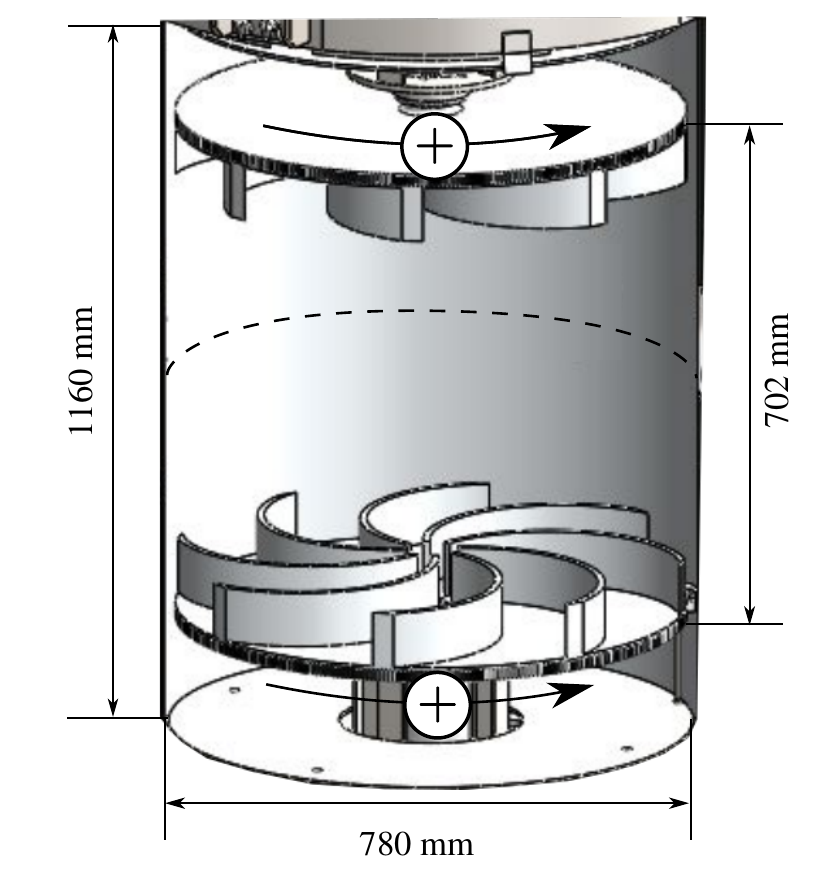}
  \caption{Schematic view of the SHREK experimental
    setup~\cite{Rousset14}. The dashed line marks the equatorial
    plane, where the sensors are located.}
  \label{fig:SHREK}
\end{figure}

The SHREK facility~\cite{Rousset14}, see \Fig{fig:SHREK}, is a
superfluid implementation of  the Von Karman flow in a
cylindrical container of inner diameter $R_s = \SI{39}{cm}$ with two
propellers of diameter $R = \SI{38}{cm}$ equipped with blades. The
distance between the turbine base disks is $h\approx \SI{70}{cm}$.

The rotation frequencies $f_1$ and $f_2$ of the bottom and top turbine
respectively can be varied independently in the range
\num{0}-\SI{2}{Hz}, which allows to produce a variety of flows from
the counter-rotating case $(f_1\times f_2 < 0 )$ to the  co-rotating
flow $(f_1\times f_2 > 0)$. See Fig.~\ref{fig:SHREK} for the $+$ rotation direction.

In the present paper, we focus on two kinds of flows:
(i) the co-rotating flow that has the bottom and the top propellers
rotating in the same $+$ direction at $f_1 = f2$,
(ii) the counter-rotating flow that has $f_1> 0$ and $-2.5 <
f_1/f_2 \leq -1.1$.

For both flows we explore three different temperatures: $T = \SI{2.3}{K}$,
\SI{2.0}{K} and \SI{1.6}{K}.

In order to operate hot-wires, the pressure
$P=\num{2.5}\pm\SI{0.1}{bar}$ is maintained above the critical
pressure.

\subsection{The probes}

Here, we describe the two sensors  that are used to derive turbulent
energy spectra: the pitot tube and the hot-wire. Both are placed in
the equatorial plane (see dashed line in \Fig{fig:SHREK}) at about
\SI{4}{cm} from the wall.

The sensors are oriented in the azimuthal direction, targeting
measurements of the  $\theta$-component of the velocity. However, it
is likely that both sensors are also sensitive to the $z$-component of
the velocity.

The acquisition frequency is nominally \SI{30}{kHz}, and data sets are
acquired over times of the order of $10^4$ large eddy turnover times,
allowing for a good statistical convergence.

We would like to emphasize that these two types of sensors were
originally proposed for measuring the velocity  in classical fluids,
mostly at room temperature.
Using them in cryogenic conditions, even in the normal fluid,
poses new challenges. This is even more problematic in the superfluid
regimes. However, using the two types of sensors simultaneously gives
a degree of confidence about the consistency of the results at least
in the large-scale range, where the normal and superfluid components
motions are  mostly synchronized by the mutual friction.

In the present paper, we have chosen not to present the results
obtained with yet another probe, a cantilever, because this probe was
located at a different distance from the wall with potentially different flow
properties.

\subsubsection{The Hot-wire}
The hot-wire is prepared from a commercial so-called
``Wollaston wire'' (see \RRef{Duri15} for details). The sensitive
part,  made of a 90\%~Platinum
10\%~Rhodium alloy,  is \SI{1.3}{\mu m} in diameter and \SI{300}{\mu m}
in length. It is etched by  electro-erosion in a 35\% nitric
acid solution. The whole wire is soldered on a DANTEC 55P01 hot-wire support.

We operate the sensor using a commercial DISA 55-M10 constant temperature
anemometer. This allows us to monitor the power needed to overheat the wire at a
fixed temperature $T_w\approx \SI{25}{K}$.

In He~I, as in standard fluids, the measurement principle is based on the
enhancement of heat transfer with forced convection. The velocity
fluctuations at length scales larger than the length of the wire can be
directly deduced from the power signal, by means of a standard King's
calibration law:
\begin{equation}
  \label{eq:king}
  e^2 = a + b v^{1/2}
\end{equation}
where $e$ is the anemometer voltage and $v$ is the velocity of the
liquid He~I flowing around the wire.

On the other hand, in He~II, the interpretation of the power signal is
trickier. The efficiency of the heat transfer is also enhanced
by forced convection and the large scale velocity
fluctuations, at small frequency, can still be deduced from the
signal~\cite{Duri15}. At higher frequency though, the signal is
marked by a spectral bump which cannot directly be attributed to
velocity fluctuations in the flow but rather to short-lived intense cooling
events, called ``glitches'', lasting typically less than a milli-second~\cite{Diribarne21}.

Since the hot-wire temperature is larger than $T_\lambda$, it is
surrounded by a thin boundary layer of He~I. Actually it is the
presence of this He~I layer that allows for the sensitivity to
velocity~\cite{Diribarne21}. Out of this layer, in He~II, the heat
flux drives an intense counterflow which, in turn, generates
additional small-scale turbulence in the form of a dense tangle of
quantized vortex lines.

\subsubsection{The Pitot tube}
In classical fluid, the Pitot tube  gives access to the dynamic pressure $s(t) = \rho
v(t)^2/2$, where $\rho$ is the density of the liquid and $v$ is velocity,
 by measuring the  pressure difference between the stagnation
pressure, at the nozzle  facing the  flow, and the static pressure at
an opening perpendicular to the flow (see \RRef{Salort21} for technical details).
Below the superfluid transition, this sensing principle remains valid at flow scales
resolved by the present sensor, because the superfluid and the normal fluid have a
common velocity at these scales. A new ``all sensor and no neck'' design is
used~\cite{Berberig98,Moukharski18}, increasing the mechanical resonance of
the sensor to about \SI{500}{Hz}. This upper frequency resolution could be
further but it would be at the expense of sacrificing the sensitivity.
The readout was capacitive and cross-band spectral averaging~\cite{Moukharski17}
was implemented.

\subsection{Flow properties}
In this section we first describe the topology of the two flow configurations
that we used, namely the co-rotating and counter-rotating flows. Then the hot-wire
measurements performed in He~I at \SI{2.3}{K} are used to assess the integral length
scale and turbulence intensity in both configurations.

\subsubsection{Topology}
\begin{figure}[htb]
  \includegraphics[width=8.5cm]{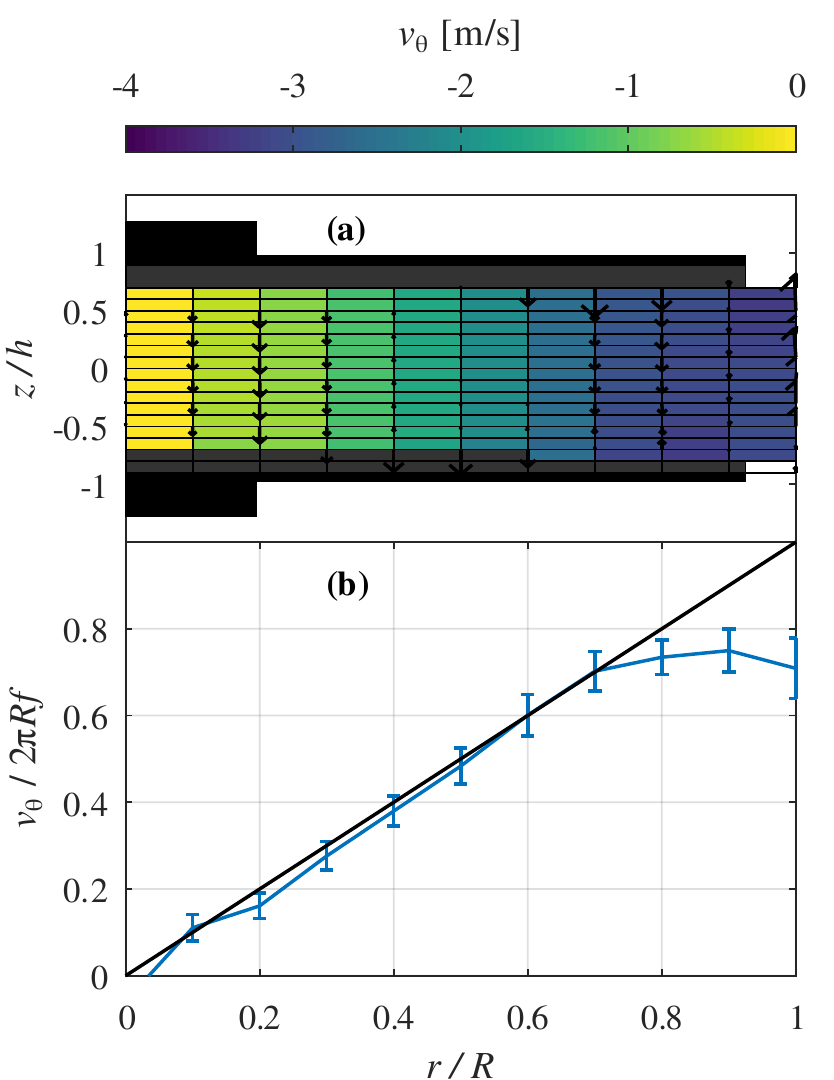}
  \caption{Azimuthal velocity $v_\theta$ of the co-rotation flow obtained in the
    SPHYNX water experiment using LDV measurements: (a) Color map of
    $v_\theta$ in the $(r,z)$ plane. Black and gray areas represent
    the turbines disk and blades respectively. Arrows indicate the
    amplitude of  the vertical velocity $v_z$, of which the maximum is
    approximately 4\% of $v_\theta (r/R = 1)$. (b):
    $v_\theta$ normalized by the velocity at the tip of the turbines
    ($2\pi Rf$) averaged  over the height of the flow. The error bars
    show the standard deviation of the azimuthal velocity.}
  \label{fig:topo_vmean}
\end{figure}

Prior to any measurements in Helium, we have explored the flow
topology and properties in a scale 1:4 experiment (denoted SPHYNX
hereafter), filled with water, using a two components Laser Doppler
Velocimetry (LDV) apparatus.
The mean $z$ and $\theta$ components of the velocity measured in water
are shown in \Fig{fig:topo_vmean}. The radial $v_r$ component (not shown) is
deduced from the other two components using the
incompressibility condition. Besides a large scale global rotation, in
the direction of the impeller rotation, one also observes a vertical
circulation, resulting from the blades curvature that induce a
pumping. The vertical circulation is  descending in the core of the
cylinder, and ascending (by incompressibility) at the wall, resulting
in a inhomogeneous large vertical shear. In the region where the hot-wire
and the Pitot measurements are performed, at $r/R\approx 0.9$, the
ratio between the azimuthal and vertical components is
$v_{z}/v_\theta\leq 4\%$.\

In the counter-rotation case,  the flow is divided into
two toric cells separated by an azimuthal shear layer, in which the mean azimuthal
velocities are zero. The position of the shear layer depends on the ratio
$|f_1|/|f_2|$: it is at equidistance from the two impellers if $|f_1|=|f_2|$,
and shifted upwards (respectively downwards) if $|f_1|>|f_2|$ (resp. $|f_1|<|f_2|$) \cite{Ravelet05,Monchaux07,Cortet10,SaintMichel14,Thalabard_2015}. Therefore,
in this paper, we only explore situations where the
rotation frequencies of the impellers are shifted ($|f_1|>|f_2|$) to
make sure that the average $\theta$ component of the
velocity is non null. Otherwise the interpretation of the Pitot and
hot-wire signals would not be possible.

\begin{figure}[ht!]
  \includegraphics[]{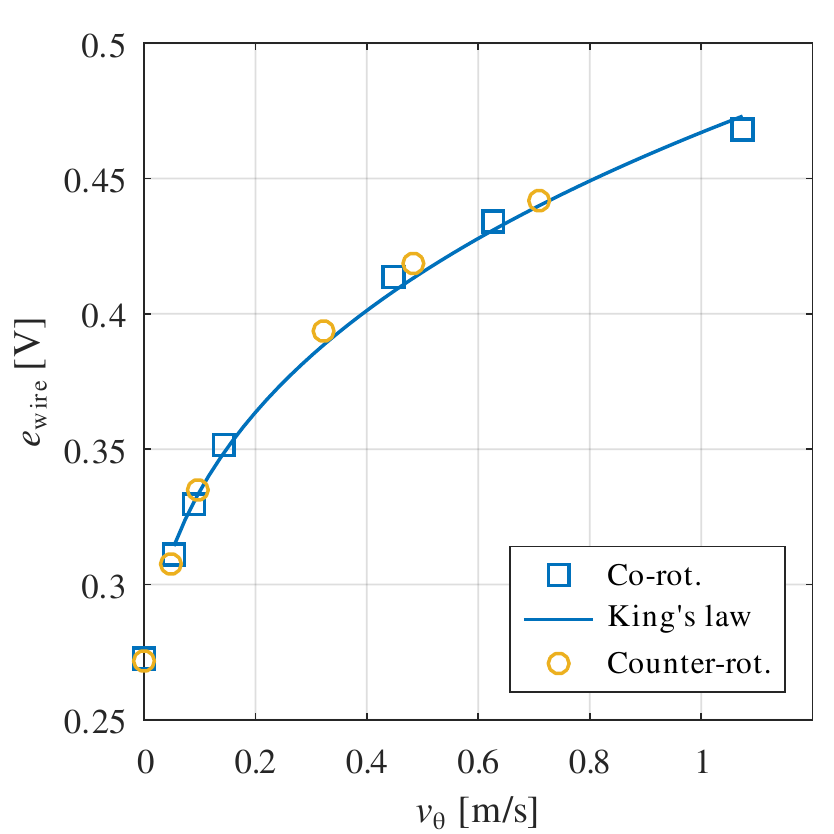}
  \caption{Mean voltage of the hot-wire anemometer as a function of the
    velocity $v_{\theta}$ defined as $v_{\theta} = \alpha 2\pi R f_1$
    where $\alpha = 0.75$ in the co-rotation case and $\alpha = 0.45$ in the
    counter-rotation case. The solid line is a fit of the co-rotation
    data  using the King's law, see Eq.~(\ref{eq:king}).}
  \label{fig:calib_hotwire}
\end{figure}

\Fig{fig:calib_hotwire} shows the calibration of the hot-wire voltage
$e_\text{wire}$  in the co-rotation and counter-rotation cases. In
absence of a reference velocity measurement in SHREK, we assumed in both cases
that the azimuthal velocity was of the form
\begin{equation}
  \label{eq:vtheta}
  v_{\theta} = \alpha 2\pi R f_1.
\end{equation}
In co-rotation, previous measurement in SPHYNX [see
\Fig{fig:topo_vmean}-(b)] suggest that using $\alpha \approx 0.75$ is
a reasonable assumption. We thus choose to take $\alpha = 0.75$
for the co-rotation case and search the value of $\alpha$ in the
counter-rotating case that leads to the best match of the mean
hot-wire voltage for a given velocity. We
find that, in counter-rotation, $\alpha \approx 0.45$, i.e. that the
velocity at the sensors location is 45\% the velocity at the tip of
the fastest turbine.

\subsubsection{Turbulence properties}
The turbulence properties of the flow are estimated both in the SPHYNX
experiment using LDV and in SHREK using the hot wire measurements
in He~I.

\paragraph{Turbulence intensity:} The turbulence intensity $\tau$ defined as
the ratio
\begin{equation*}
  \tau = \sigma_v/|{\bf v}|,
\end{equation*}
where ${\bf v}=v_\theta {\bf e}_\theta+v_z {\bf e}_z$ and $\sigma_v =
\sqrt{\langle v'^2\rangle}$
is the standard deviation of the module of the velocity ${\bf v}$.
At a distance of order \SI{4}{cm} from the wall, i.e. at
coordinate $r/R\approx 0.9$ in \Fig{fig:topo_vmean}-(b), the turbulence
intensity is found to be in  the range 5--10\%. This order of
magnitude is confirmed by hot-wire measurements in co-rotating He~I,
where the inferred value is $\tau\approx 5.2\%$~\cite{Salort21}.
Using the same technique, and the calibration from
\Fig{fig:calib_hotwire} one finds $\tau\approx 22\%$ in
counter-rotation, i.e. a turbulence intensity which is 4-5 times larger
than in co-rotation.

\paragraph{Integral length scale:} We used the hot-wire velocity signal
to compute the longitudinal integral length scale $L_l$ defined as
\begin{equation*}
  \label{eq:L}
  L_l = \int_0^{+\infty} \frac{<v'(0)v'(r)>}{\langle v'^2 \rangle}d(\delta r)
\end{equation*}
As shown in \RRef{Salort21} this leads to $L_l\approx \SI{2.9}{cm}$ in
co-rotation, while we find $L_l\approx \SI{3.7}{cm}$ in the
counter-rotation case.

\section{Local energy spectra}
\label{sec:spectra}
In this section we present power spectral density of the hot wire
signal in both co- and counter-rotating flows.

Since the large scale behavior of those flows is not expected to be
affected by the transition to superfluid phase~\cite{Maurer98,SaintMichel14},
we first present measurements in He~I where the hot-wire is expected to behave as a
standard  anemometer. Those spectra are further used as references
and compared to those obtained in He~II.

\subsection{Normal Fluid}

\begin{figure}[htb]
  \includegraphics[]{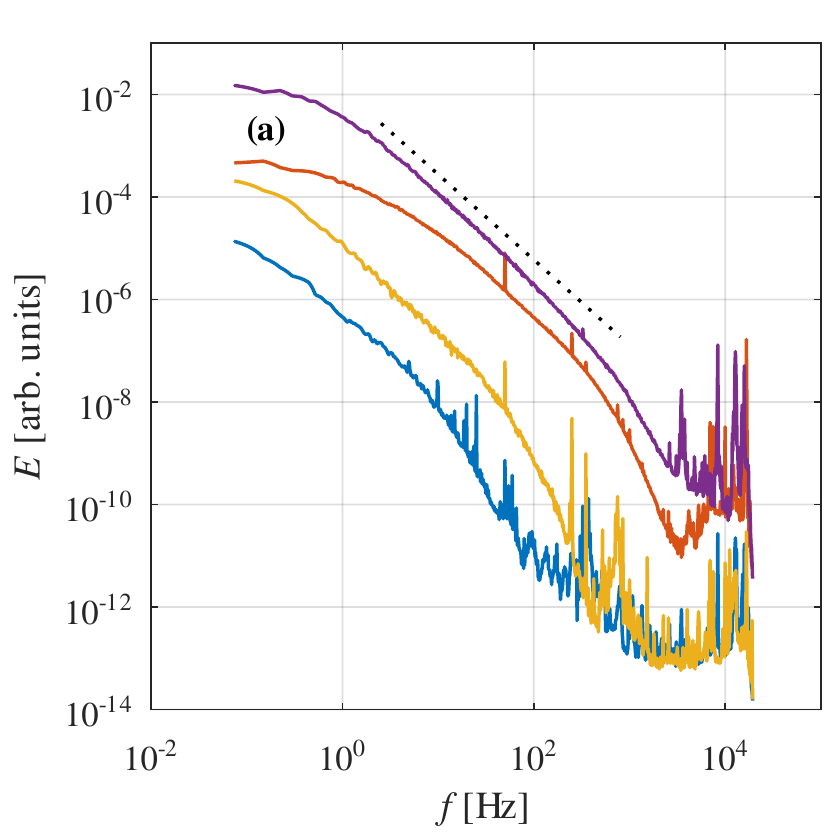}
  \includegraphics[]{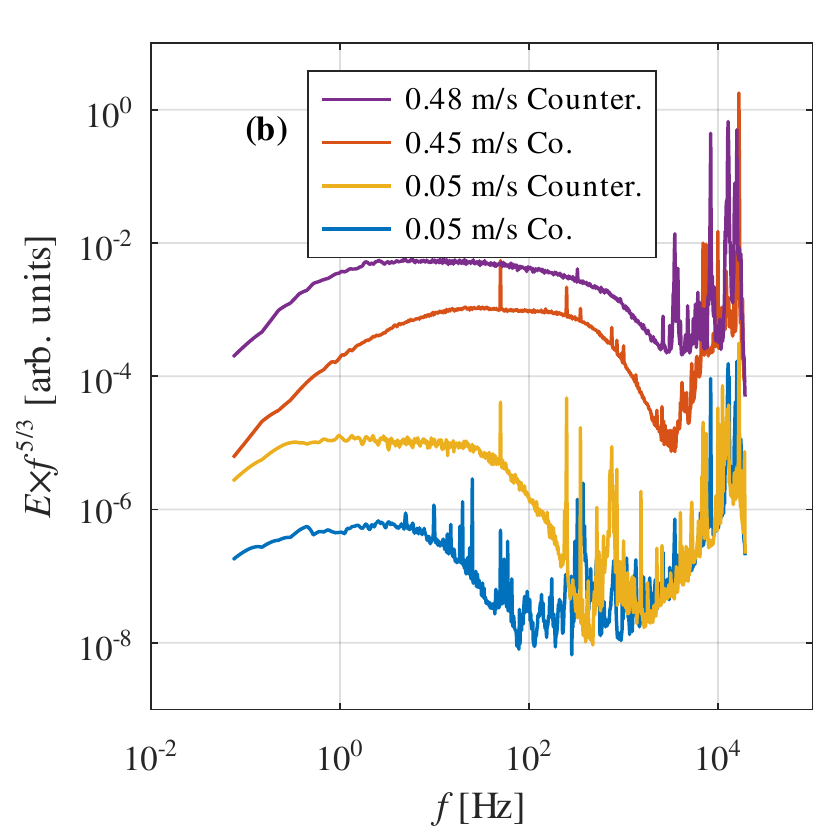}
  \caption{(a): Power spectral density of the hot-wire signal in He~I at
    \SI{2.3}{K}. Amplitudes are shifted arbitrarily for better readability.
    The dotted  lines show a $f^{-5/3}$ power law.
    (b): Same spectra compensated by $f^{5/3}$.}
  \label{fig:spectra_heI}
\end{figure}

\Fig{fig:spectra_heI}-(a) shows the power spectral density (PSD
hereafter) of the
hot-wire signal in co-rotation and in counter-rotation at two
comparable azimuthal velocities. In order to make sense out of those
spectra we assume the Taylor hypothesis of frozen turbulence, so that
we can translate a given frequency $f$ to a length scale $l$ through the
relation $l = \langle v_\theta\rangle / f$. Note though that this
hypothesis is probably not justified in the case of counter-rotation,
where the turbulence intensity is very high, but this should only
matter at the highest frequencies.

The spectra are flat at low frequency and then tend to follow a power
law at higher frequency, where the inertial range of length scales
is expected to lie. At even higher frequency a cut-off if
observed. The compensated spectra in
\Fig{fig:spectra_heI}-(b) show that the power-law in the inertial
range is compatible with a Kolmogorov $f^{-5/3}$ energy cascade in
both flows. The transition from the low frequency uncorrelated flat
spectrum to the power law is quite different in co-rotation and in
counter-rotation though. Since the integral length scales are
comparable in both flows, we expect that the transition happens
at comparable frequency for a given azimuthal velocity. Even though
the transition from flat to power law behavior actually seems to
happen at comparable frequencies, in counter-rotation it is much more
steep than in co-rotation where the slope evolves gradually from 0 to
$-5/3$ over a decade of frequencies.

The interpretation of the cut-off at large frequency calls for
caution. At low velocity, it happens at lower frequency in
the co-rotating than in the counter-rotating case. If the cut-off
marks the beginning of the dissipative length scales, this is expected
since the turbulence intensity of the latter is much higher than the
former.
At high velocity though, we can hardly distinguish the cut-off
frequencies and it is likely that it should be attributed to a
finite size effect.

\subsection{Superfluid}

\begin{figure}[htb]
  \includegraphics[]{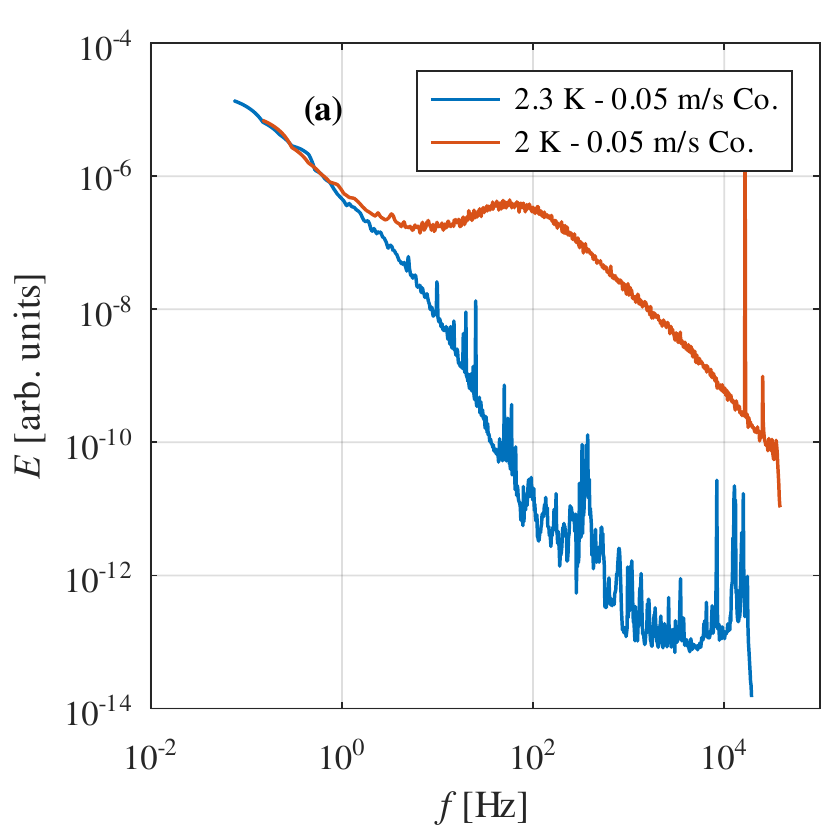}
  \includegraphics[]{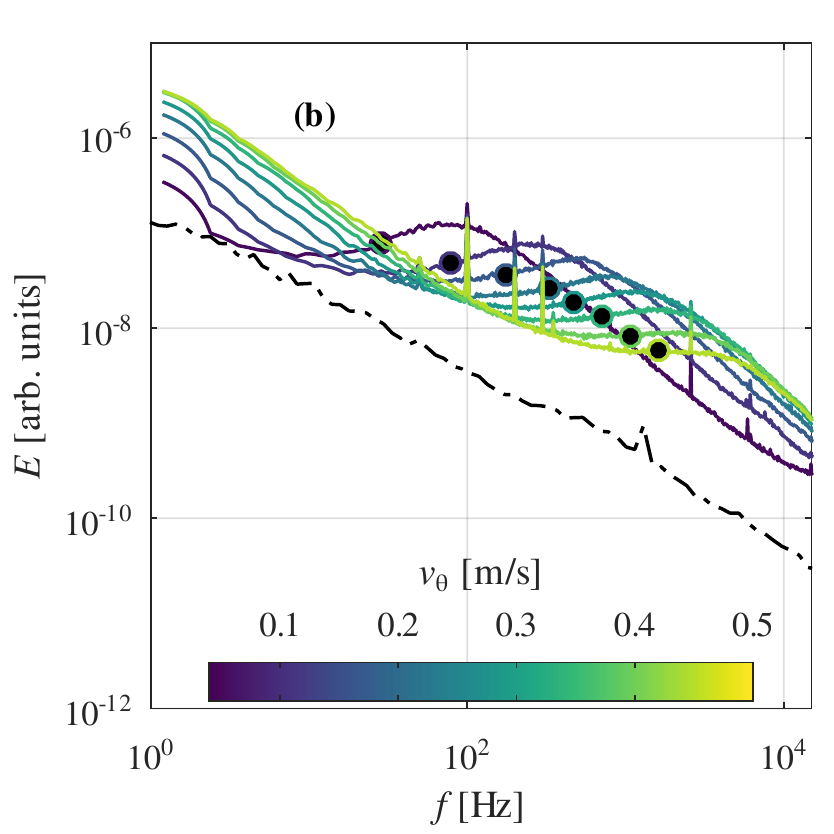}
  \caption{(a): Power spectral density of the hot-wire signal in He~I
    (\SI{2.3}{K})  and in He~II (\SI{2}{K}) in co-rotation at
    \SI{0.07}{m/s}. Amplitudes are shifted so that the spectra match
    at low frequency.
    (b): Series of spectra in He~II (\SI{2}{K}) in co-rotation at
    velocities varying in the range \num{0.04}--\SI{0.45}{m/s}
    (colored solid lines) and at \SI{0}{m/s} (black dash dotted). The
    black dots mark the inflection point in the high frequency bump.}
  \label{fig:spectra_heII}
\end{figure}

\Fig{fig:spectra_heII}-(a) compares the PSD of the hot-wire raw signal
in He~I (\SI{2.3}{K})  and in He~II (\SI{2}{K}) in co-rotation at low
velocity. In He~II, we see that a large spectral bump appears at high
frequencies, where, in He~I, the PSD is already damped by the viscous
cutoff. This spectral bump is actually associated with short-lived heat flux
enhancement events that account for a significant, velocity
dependent, portion of the variance of the hot-wire signal.
Thus velocity  fluctuations cannot be directly inferred from the
hot-wire raw signal.

In \Fig{fig:spectra_heII}-(b) we show PSD obtained in He~II
(\SI{2}{K}) at increasing azimuthal velocities, in co-rotation flow.
It is clear that the frequency at which the spectral bump appears
increases with the flow velocity. In quiescent helium, no bump is
observed, down to the lowest resolved frequencies.
Those features have also been reported
in Ref.~\cite{Diribarne21} but within a more limited range of velocities
and in a grid flow where the turbulence  intensity is very low
(less than 2\%). Note that, contrary to previous observations, while at
low velocity a local maximum is observed, at high velocity the bump takes
the form of departure from the low frequency power law behavior with no clear
extremum.

\subsection{Comparison with Pitot - velocity spectra}
While the Pitot tube has a lower spatial resolution, the
interpretation of its signal is more straightforward. Especially in the
case of the co-rotation flow, where the turbulence intensity is
low, the Pitot signal fluctuations can be shown to be  linearly
related to velocity fluctuations in the flow.

\begin{figure}[ht!]
  \includegraphics[width=8.5cm]{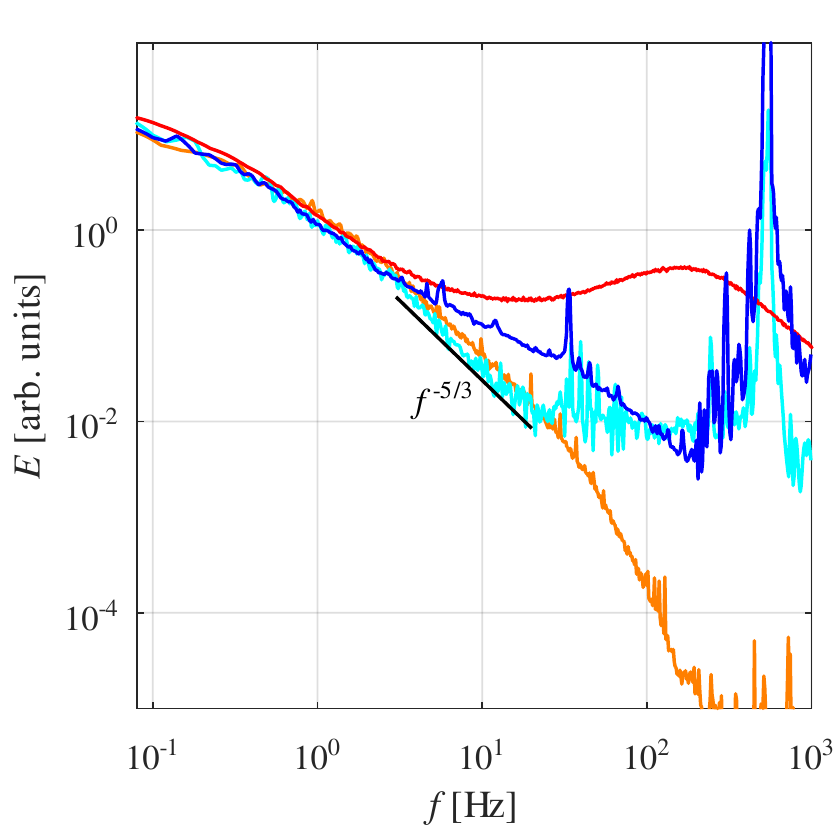}
  \caption{Comparison of the Pitot and hot-wire signal spectra, $E$,
  at \SI{2.3}{K} (cyan and orange resp.) and \SI{2}{K} (blue
  and red resp.)  in co-rotation at \SI{0.09}{m/s}.  The  solid black
  lines shows a $f^{-5/3}$ power law. 
  An arbitrary scaling factor is applied so that the amplitudes match
  at \SI{1}{Hz}.}
  \label{fig:compare_pit_hw}
\end{figure}

\Fig{fig:compare_pit_hw} shows a comparison of the Pitot and hot-wire
signal PSD in co-rotation. In He~I, the shape of the  PSD of
the two sensors are very similar  at low frequencies:  after a
non-universal shallower spectrum at  low frequencies, the PSD shows
a $f^{-5/3}$ power law from  $f\approx\SI{5}{Hz}$ up to $f\approx \SI{20}{Hz}$
where the spectrum reaches a noise plateau. The latter can be
explained by the low sensitivity of the Pitot sensor at low
velocity. The peak in the Pitot spectrum at $f\approx \SI{540}{Hz}$ is
due to the probe mechanical resonance.

As shown by \textcite{Salort21}, in He~II at the same velocity, the PSD of the
Pitot remains unchanged up to $f\approx \SI{3}{Hz}$ where a departure
is observed: instead of tending to a  $f^{-5/3}$ power law like in
He~I, the PSD amplitude keeps decreasing like $\sim f^{-1}$ until it
reaches the tail of the probe mechanical peak, at $f\approx \SI{200}{Hz}$.

A departure from the He~I PSD is also observed at approximately the same
frequency (around \SI{3}{Hz}) but it is not as pronounced as for the hot-wire.

This departure in the Pitot spectrum is attributed to the pile-up of
kinetic energy in the superfluid component in the near dissipative
range of length scales~\cite{Salort11b,Salort21}.

\section{Discussion}
\label{sec:discussion}
In this section we will to try explain the shape of the hot-wire
spectra and underpin the origin of the high frequency bump in the hot-wire signal.

\subsection{Velocity dependence}
In order to analyze the velocity dependence of the high frequency bump
in the hot-wire signal, we define the
representative frequency $f_\text{bump}$ as the local inflection point
between the low frequency power law and the high frequency spectral
departure [see the  black dots in \Fig{fig:spectra_heII}-(b)].
Contrary to previous studies, the bump here does not
always feature a maximum, and this definition guaranties that we can
always find a representative frequency for the bump.
Qualitatively, $f_\text{bump}$ can be viewed as the lowest frequency at which
the bump starts.

The inflection point is located automatically  by first
fitting the PSD at intermediate frequencies with a third order
polynomial and looking for a local maximum in the derivative $dE/df$.

\begin{figure}[ht!]
  \includegraphics[width=8.5cm]{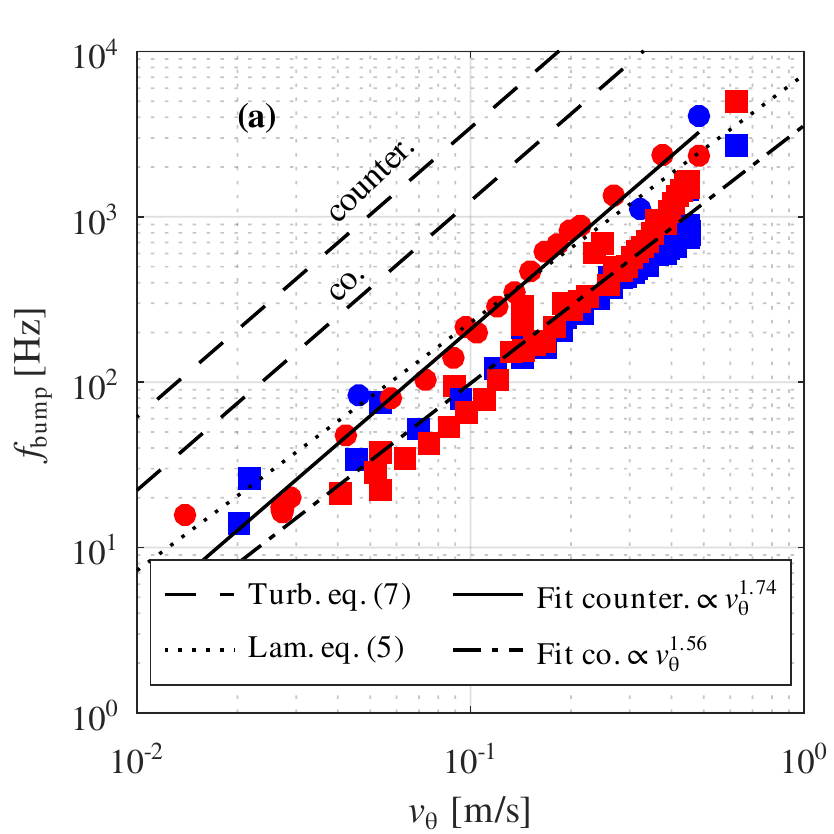}
  \includegraphics[width=8.5cm]{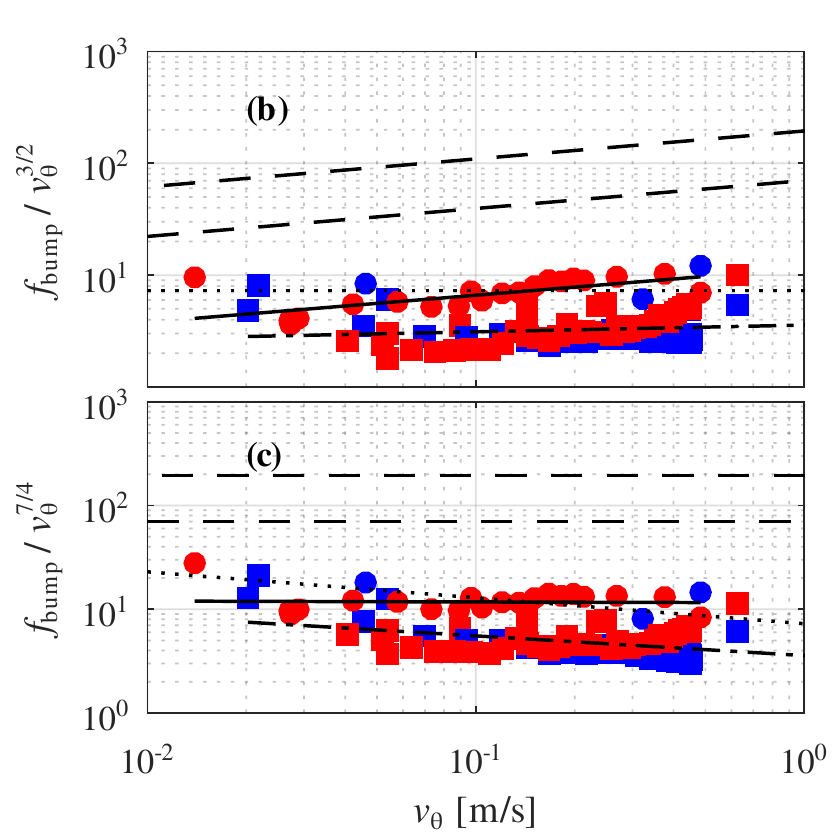}
  \caption{Representative frequency of the spectral bump
    $f_\text{bump}$ as a function of the
    azimuthal velocity $v_\theta$ in counter-rotation (square symbols)
    and co-rotation (round symbols) at \SI{2}{K} (red) and \SI{1.6}{K}
    (blue). The solid and dash-dotted black lines show the best
    fit of the form $f_\text{bump}\propto v_\theta^\gamma$ for the
    counter-rotating and co-rotating cases respectively. The dashed
    lines correspond to eq.~(\ref{eq:7/4}) computed for both kinds of
    flows, while the dotted line corresponds to eq.~(\ref{eq:3/2}).
    (a): raw frequency. (b): frequency divided by $v_\theta^{3/2}$.
    (c): frequency divided by $v_\theta^{7/4}$. }
  \label{fig:bump_freq}
\end{figure}

\Fig{fig:bump_freq}(a) shows $f_\text{bump}$ as a function of the
azimuthal velocity $v_\theta$ for both co-rotation and
counter-rotation cases at \SI{2}{K} and \SI{1.6}{K}.
The representative frequency is extracted either from experiments at
steady or at very slowly ($\approx\SI{1e-5}{Hz.s^{-1}}$) varying
turbine frequency. In the case of varying frequency, each point is
extracted from a spectrum averaged over ten
consecutive datasets lasting $\approx \SI{54}{s}$ each (about 800 integral
times at the smallest rotation frequency). Only points
for which the turbine frequency varies of 15\% at most
between the first and the last dataset are shown.

From this figure, one already notices two striking features:
\begin{itemize}
  \item the bump appears at higher frequency in counter-rotation
  (round markers) than in co-rotation (square markers) for a given
  azimuthal velocity,
  \item the bump frequency does not significantly depend on the
  temperature (red versus blue markers).
\end{itemize}

The bump is not visible in quiescent fluid at the resolved frequencies
(see Fig.\ \ref{fig:spectra_heII}(b)).
It is therefore reasonable to assume that $f_\text{bump}$
tends to \SI{0}{Hz} when the velocity tends to \SI{0}{m/s}. The
investigation of the emergence of the bump in the low velocity limit
would require a dedicated 
campaign with very large acquisition times, and is beyond the scope
of this paper. In the range of velocities investigated here, the velocity dependence 
of the bump frequency can be represented as a simple
power law. The solid and dash-dotted black lines in
\Fig{fig:bump_freq} indicate,  
in counter-rotation and in co-rotation respectively, the best fits of
the form
\begin{equation}
f_\text{bump} \propto v_\theta^\gamma.
\end{equation}

The exponent $\gamma$ is higher in counter-rotation ($\gamma = 1.74$)
than in co-rotation ($\gamma = 1.56$). Note
that since the estimated azimuthal velocity in counter-rotation is
calibrated against that in co-rotation, the observed difference cannot
be attributed to a wrong value for $\alpha$ in eq.~(\ref{eq:vtheta}).
Anyway, the velocity range here is about 1.5 decades, much larger than
in previous studies~\cite{Diribarne21}, which strongly supports the view that the bump frequency
dependence with the velocity is steeper than a simple linear dependence.

We identify below some of the relevant characteristic
frequencies that can emerge in a rotating turbulent flow and we detail
their respective velocity dependence.

\paragraph{Vortex-streets emanating from the wire:}
\textcite{Diribarne21} have shown that the normal and superfluid
components form two well defined ``wing-like'' patterns in the vicinity
of the wire. The characteristic size of the patterns,
was shown to be
typically hundred times the diameter of the wire in their working
conditions. They further argue that this flow pattern should be
unstable and could lead to K\'arm\'an vortex streets in the wake of the ``wing''.
Assuming the hot-wire heat  flux is affected by this vortex shedding,
this would lead to a frequency:
\begin{equation}
  \label{eq:karman}
  f_\text{K\'arm\'an} = \frac{2 \text{St} v_\theta}{D(T, v_\theta)} ,
\end{equation}
where $\text{St}$ is the Strouhal number~\cite{strouhal1878} of the
order 0.1 -- 0.3, and $D(T, v_\theta)$ is the temperature and velocity dependent
characteristic size of the thermal wing pattern. The dependence of
$D$ on the velocity has been shown to be of the order $\propto
v_\theta^{-1}$ in cylindrical approximation, and to tend towards
$\propto v_\theta^{-1/2}$ when $D$ becomes large as compared to the
length of the wire. It thus predicts a vortex
shedding frequency $f_\text{K\'arm\'an} \propto v_\theta^\beta$ with
$1.5\lesssim \beta\lesssim 2$.

 \paragraph{Frequency corresponding to intervortex distance:}
Because turbulence in SHREK is inhomogeneous, the intervortex
distance is expected to vary depending on the position in the flow.
We can derive two limiting formulae for the frequency corresponding to
the intervortex distance $f_\delta $, assuming that   we are in a
co-rotating laminar flow or a fully turbulent regime.
In the first case,  we can take as a reference  the distance between
the vortex neighbors in a laminar superfluid uniformly rotating with
frequency $f_{r}$, which is likely to be the lower bound since the
mean vorticity of the turbulent flow is larger
than the one in the laminar flow. In this case, the vortex line density
is given by $\mathcal{L} = 4\pi f_r/\kappa$ (see, e.g.,
\Refs{Feynman55,Yarmchuk79}) and the intervortex distance
by $\delta = \mathcal{L}^{-1/2} = (4\pi f_r/\kappa)^{-1/2}$.
Assuming that the vortex array is advected at the same velocity as the
mean flow, we consequently find that the typical frequency $f_\delta^{lam}$
corresponding to such a reference scale is $v_\theta/\delta$, hence:
\begin{equation}
\label{eq:3/2}
f_\delta^{lam} =  \left(\frac{2}{\kappa R}\right)^{1/2} v_\theta^{3/2}.
\end{equation}

Let us now  consider turbulence when estimating the
intervortex scale. It has been shown that in the hypothesis of
homogeneous and isotropic turbulence (HIT hereafter), the intervortex
spacing scales like the Kolmogorov dissipative length scale~\cite{Salort11b,Babuin14}:
\begin{equation}
  \label{eq:babuin14}
  \frac{\delta}{L_l} = \left(\frac{\nu_\text{eff}}{\kappa}\right)^{1/4}
                     \text{Re}^{-3/4}_\kappa,
\end{equation}
where $\nu_\text{eff}$ is determined experimentally (see
e.g. Refs.~\cite{Smith93,Stalp99,Babuin14}) and
$\text{Re}_\kappa = \sigma_v L_l/\kappa$ is the
turbulent Reynolds number. Using the Taylor
hypothesis, Eq.~(\ref{eq:babuin14}) translates to a frequency in the
Eulerian frame
\begin{equation}
  \label{eq:7/4}
  f_\delta^{turb} = \left(\frac{\tau^3}{\nu_\text{eff} L \kappa^2}\right)^{1/4}
  v_\theta^{7/4}.
\end{equation}

The vortex shedding model [Eq.~(\ref{eq:karman})], predicts a velocity
dependence of the shedding frequency compatible with the data for
$f_\text{bump}$. On the other hand, in this basic model, the amplitude
of the velocity fluctuations relative to the mean velocity, i.e. the
turbulence intensity, do not play any role and this is in
contradiction with the fact $f_\text{bump}$ is found to have notably
different values in the co-rotating and counter-rotating situations
for a given mean velocity. Additionally, we expect that in this model,
including some fluctuations around the mean velocity would probably
increase the standard deviation of the shedding frequency rather than
changing its mean value. Moreover, it was shown~\cite{Diribarne21}
that due to the temperature dependence of the characteristic thermal pattern
size $D$ in Eq.~(\ref{eq:karman}), the spectral bump frequency should depend
noticeably on the temperature. For those reasons, the shedding model, in its
current basic form, seems unable to account for the present measurements.

The frequency associated to the intervortex distance, is expected to
scale as $v_\theta^{3/2}$ or $v_\theta^{7/4}$ for the laminar and
turbulent cases respectively. The compensated plots
\Fig{fig:bump_freq}(b) and \Fig{fig:bump_freq}(c) show that both
exponents are are good candidates, even though counter-rotation data seem
to have a slightly steeper slope, as seen from the
fits (solid and dashed-dotted line in \Fig{fig:bump_freq}).
The expected frequency $f_\text{bump}$ for the laminar
(dotted line in \Fig{fig:bump_freq}) and the turbulent (dashed lines
in \Fig{fig:bump_freq}) show that the estimated frequencies, are in qualitative
agreement in both cases.
The true motion is clearly neither  purely laminar nor statistically isotropic:
it consists of both (an anisotropic) turbulence and a rotational mean flow.
Therefore, the scaling of $f_\text{bump}$ should be somewhere in between of the
purely laminar and the purely turbulent scalings. However, since the
latter two scalings are very close to each other, we conclude that our
conclusion that $f_\text{bump}$ is associated with the intervortex
spacing is robust.

\begin{figure}[htb]
  \includegraphics[width=8.5cm]{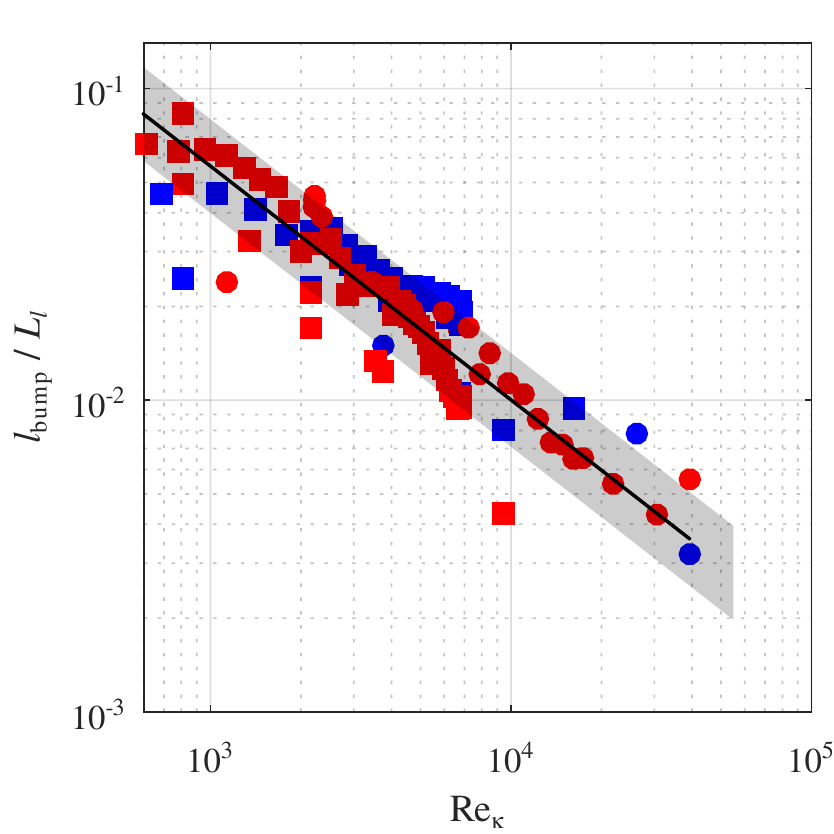}
  \caption{Characteristic length $l_\text{bump} =
    v_\theta/f_\text{bump}$ normalized by the corresponding flow
    integral length scale $L_l$ as a function of the turbulent Reynolds
    number. The black line represents Eq.~(\ref{eq:babuin14})
    multiplied by an arbitrary factor 15.}
  \label{fig:bump_length}
\end{figure}

In \Fig{fig:bump_length} we show the characteristic length
$l_\text{bump} =  v_\theta/f_\text{bump}$, normalized by the
integral length scale $L_l$, as a function of the turbulent Reynolds
number $\text{Re}_\kappa = \tau v_\theta L_l/\kappa$. This
representation collapses the data from both kinds of flows onto a
reasonably well defined single power law. For comparison, the black
line represents the intervortex distance normalized by the integral
length scale, Eq.~(\ref{eq:babuin14}), multiplied by an arbitrary
factor 15. In the range of temperatures between \SI{1.6}{K} and
\SI{2}{K}, the effective viscosity $\nu_\text{eff}$ has been
shown to not depend significantly on the temperature (see, e.g., the
compilation of experimental and numerical data from \RRef{Babuin14})
and we consequently used the average reported value
$\nu_\text{eff}\approx \kappa/5$~\cite{Babuin14}.
As a guide to the eye, the gray area shows the region around this
line into which the data are scattered by at most a factor of two.

Even though the data are still scattered, it is reasonable to
assume that we should search the origin of the spectral bump in
phenomena that are prominent at length scales proportional to the
intervortex spacing.

\subsection{Interpretation}
In the superfluid thermal boundary layer, the very intense counterflow
heat flux results in a dense vortex tangle. The intervortex
distance varies radially through the thermal (He~II) boundary layer:
the heat flux decreases as one gets further from the wire,
due to the cylindrical geometry, and so does the vortex line density.
So no single length scale can be identified in the thermal boundary layer,
but close to the wire, where the temperature gradient is significant,
the intervortex distance is orders of magnitude smaller (see \RRef{Duri15})
than that of the bulk surrounding turbulent flow.

\textcite{Diribarne21} have shown that the spectral bump is actually
the result of short-lived intense cooling events named ``glitches''.
They did not devise a mechanism by which those sudden enhancements of
the heat transfer could be triggered but envisaged two possible leads: (i)
the shedding of vortices passed the wire, (ii) the destabilisation of
the vortex tangle around the wire due to the bottlenecking, or pile-up
of kinetic energy, in the  superfluid component at scales
comparable with the intervortex distance, as predicted in \RRef{Salort11b}.

As shown in the previous section, we now have  arguments to
eliminate (i), due to the dependence of the bump frequency on the
turbulence intensity. The apparent independence of the bump frequency
on the temperature is another argument against this explanation,
as already noted in \RRef{Diribarne21}.
On the other hand, we can certainly settle on the
fact that the process triggering those glitches should occur at
small scales. Lead (ii) is appealing,
because the hot wire bump seems to happen at frequencies comparable
with those  at which a pile-up of kinetic energy happens, as measured
by the Pitot tube (see \Fig{fig:compare_pit_hw}). This is only
qualitative: due to the very limited set of velocities where
the Pitot tube has a sufficient spacial resolution to show the
pile-up, we cannot prove that there is an actual correlation with the
appearance of the spectral bump in the wire signal.

Following lead (ii), a mechanism explaining the influence of
the hot-wire signal to quantum intervortex distance in the outerflow
is as follows:
in a mechanically driven quantum turbulence,  the mutual friction
between the normal and superfluid components couples their turbulent
fluctuations: $\B u\sb n(\B r,t ) \approx  \B u\sb s(\B r,t)$  at all
scales larger than the intervortex scale $\delta$. The resulting
turbulent energy spectra of the mechanically driven quantum
turbulence for the scales much greater than $\delta$ are close to those of
the classical hydrodynamic turbulence
\cite{Maurer98, Barenghi14,Salort10,Salort11,Roche09,Barenghi14b,Eltsov14}.
However, $\B u\sb n(\B r,t ) $ and $ \B u\sb s(\B r,t )$ decouple at
scales of the order of $\delta$.
Roughly, such relative motion of the normal fluid and the superfluid
vortex tangle can be viewed as a normal flow past an irregular ``grid"
made of the quantized vortex lines. Naturally, such a flow produces
extra turbulence at the ``grid spacing" scale, i.e. at the scales
comparable to $\delta$. More precisely, on a microscopic level,
the normal fluid is a field of acoustic phonons which scatter off
the quantized vortices and thereby acquire spatial inhomogeneity with
a characteristic scale of the order of the mean distance between such
vortex scatterers.
Obviously, the energy of the bump cannot come from ``nowhere'' i.e. it
could only appear as a result of transfer from the mean relative
motion at larger  scales. A good candidate for such a mean motion is
the thermal counterflow produced by the wire. In this case, the bump
is indeed a product of the intrusive nature of the hot-wire and, at the
same time, its properties are affected by the surrounding turbulent
flow.
This is a simple and robust qualitative mechanism of the spectral bump  creation
near the intervortex scale. However, for completeness
let us mention another possible mechanism for the spectral bump generation.

A third mechanism could explain the heat flux glitches experienced by
the hot wire at frequencies corresponding to the small length scales
of the  external turbulence: The presence of intense vorticity and
pressure structures associated with bundles of quantum
vortices. Those objects are the counterpart in quantum turbulence
of ``vorticity worms'' well known in classical turbulence (see, e.g.,
the pioneering numerical and experimental works
\Refs{siggia1981numerical,DouadyFilament:PRL1991}).
The existence of vortex bundles have been reported in quantum turbulence,
both numerically~\cite{Baggaley_Coherentvortexstructures_EPL2012} and
experimentally~\cite{Rusaouen17}. Their typical associated length
scale (diameter) was reported to be around two times the intervortex
distance in
superfluid~\cite{Baggaley_Coherentvortexstructures_EPL2012} or four
times the Kolmogorov viscous length scale~\cite{jimenez1998characteristics}.
The pressure signature of superfluid vortex bundles
was measured in the SHREK apparatus~\cite{Rusaouen17}, and the
authors evidenced that, here again, no real difference could be made
between classical and quantum turbulence.
When such a vortical structure impinges the wire, we expect it to
 polarize the vortex tangle constituting the thermal boundary layer,
leading to a change in its effective thermal conductivity. Indeed,
it was shown that heat transfer can be modeled by standard counterflow
phenomenology. In this framework, the mutual friction force per unit
volume between the counter-flowing normal and superfluid components,
is the key ingredient in the definition of a local conduction function.
The latter relates the local temperature gradient $\bm{\nabla}T$
in the He~II boundary layer with the heat flux $\bm{\varphi}$ and
in some way it can be seen as an effective thermal conductivity.
A theoretical expression of the conduction function $f(T)$
can be obtained at heat fluxes well above the critical heat flux
at which the counterflow becomes turbulent~\cite{Vinen57c,Vansciver12}:
\begin{equation}
  \label{eq:FT}
  f(T) = C\frac{2\rho\rho_s^3 s^4 T^3}{\gamma^2 B \rho_n \kappa},
\end{equation}
where $f(T) = |\bm{\varphi}|^3/|\bm{\nabla}T|$ is the conduction function,
$\rho_n$ and $\rho_s$ are the normal and superfluid density respectively,
$s$ is the entropy per unit mass, $B$ is a constant of order unity
(see e.g. \RRef{Lucas70}), $\gamma$ is defined as $\mathcal{L} = \gamma^2 (\bm{v_n}-\bm{v_s})^2$ where $\mathcal{L}$ is the local vortex line density of the counterflow
(see \RRef{Tough82}), and $C$ depends on the average angle between
the vortex lines and the heat flux. For an isotropic vortex tangle, $C=3/2$,
while this constant tends towards infinity when the vortices are polarized and oriented
parallel to the counterflow velocity $(\bm{v_n}-\bm{v_s})$.
This continuous approach proved efficient in modeling heat  transfer from heat wire down to micron scales~\cite{Diribarne21, Duri15}.
Knowing the collision frequency of the vortical structures on the wire would help to
confirm or invalidate this mechanism. Although we have not been able to
find previous studies on this specific question, it seems reasonable to assume that the
typical collision time scales are linearly related to the time scales of the smallest flow structures, such as the intervortex one. This would be consistent with the scaling reported in Fig.~\ref{fig:bump_length}.

\section{Conclusions}
In this paper, we report experimental measurements in liquid helium  using
a hot-wire probe and a pitot tube. These measurements are done in the SHREK
facility in both He~I and He~II, for different levels
of co-rotation or counter-rotation. In normal fluid, we use the hot-wire to
devise the integral length scale and turbulence intensity of both flows. This
allows us to compute the turbulent Reynolds number in each case.

In He~II the hot-wire signal exhibits a spectral bump at high
frequency, of which the representative frequency increases with the
velocity $v_\theta$, as previously reported, but also with the
turbulence intensity of the flow. We show that the latter cannot be
explained satisfactorily by the model of vortex shedding as proposed in
\RRef{Diribarne21}.

The velocity dependence is compatible with a power law $v_\theta^\gamma$
over more than one decade of frequencies,
with $\gamma$ in the range $1.5\lesssim\gamma\lesssim 1.8$.
Assuming that the frequency of the quantum bump
can be translated to a length scale of the flow by use of the Taylor hypothesis,
we have presented the resulting length $l_\text{bump}$ as a function of the
turbulent Reynolds number. This representation collapses data from both
co-rotating and counter-rotating flows onto a single power law compatible with
$l_\text{bump} \propto \delta$.

Thus the phenomenon that triggers the quantum bump must happen at
scales proportional to the intervortex distance. We recall that the
spectral bump is actually the
result of thermal "glitches", short lived heat transfer improvement events,
in the time domain. We propose two possible qualitative scenarios that end up
destabilizing the  wire's thermal boundary layer, leading to fluctuations of
its overall thermal resistance:
\begin{itemize}
    \item the interaction between the wire's counterflow and the
    enhanced velocity fluctuations of the flow,
    \item the polarization of the vortex tangle of the wire by the
    vortical structures associated to turbulence.
\end{itemize}

Those explanations are of course qualitative, and some further numerical
and experimental studies are needed to understand the quantitative
aspects of the quantum bump generation.
\section{Acknowledgements}
This work has been supported by the French Agence National de la
Recherche (SHREK, grant no. ANR-09-BLAN-0094-01), and  by the European High-performance Infrastructures in Turbulence (CE-EuHIT, project “Felisia”, grant no. 312778).

\bibliography{SHREK_HotWire.bib}

\begin{thebibliography}{39}%
\makeatletter
\providecommand \@ifxundefined [1]{%
 \@ifx{#1\undefined}
}%
\providecommand \@ifnum [1]{%
 \ifnum #1\expandafter \@firstoftwo
 \else \expandafter \@secondoftwo
 \fi
}%
\providecommand \@ifx [1]{%
 \ifx #1\expandafter \@firstoftwo
 \else \expandafter \@secondoftwo
 \fi
}%
\providecommand \natexlab [1]{#1}%
\providecommand \enquote  [1]{``#1''}%
\providecommand \bibnamefont  [1]{#1}%
\providecommand \bibfnamefont [1]{#1}%
\providecommand \citenamefont [1]{#1}%
\providecommand \href@noop [0]{\@secondoftwo}%
\providecommand \href [0]{\begingroup \@sanitize@url \@href}%
\providecommand \@href[1]{\@@startlink{#1}\@@href}%
\providecommand \@@href[1]{\endgroup#1\@@endlink}%
\providecommand \@sanitize@url [0]{\catcode `\\12\catcode `\$12\catcode
  `\&12\catcode `\#12\catcode `\^12\catcode `\_12\catcode `\%12\relax}%
\providecommand \@@startlink[1]{}%
\providecommand \@@endlink[0]{}%
\providecommand \url  [0]{\begingroup\@sanitize@url \@url }%
\providecommand \@url [1]{\endgroup\@href {#1}{\urlprefix }}%
\providecommand \urlprefix  [0]{URL }%
\providecommand \Eprint [0]{\href }%
\providecommand \doibase [0]{https://doi.org/}%
\providecommand \selectlanguage [0]{\@gobble}%
\providecommand \bibinfo  [0]{\@secondoftwo}%
\providecommand \bibfield  [0]{\@secondoftwo}%
\providecommand \translation [1]{[#1]}%
\providecommand \BibitemOpen [0]{}%
\providecommand \bibitemStop [0]{}%
\providecommand \bibitemNoStop [0]{.\EOS\space}%
\providecommand \EOS [0]{\spacefactor3000\relax}%
\providecommand \BibitemShut  [1]{\csname bibitem#1\endcsname}%
\let\auto@bib@innerbib\@empty
%</preamble>
\bibitem [{\citenamefont {Diribarne}\ \emph {et~al.}(2021)\citenamefont
  {Diribarne}, \citenamefont {Rousset}, \citenamefont {Sergeev}, \citenamefont
  {Valentin},\ and\ \citenamefont {Roche}}]{Diribarne21}%
  \BibitemOpen
  \bibfield  {author} {\bibinfo {author} {\bibfnamefont {P.}~\bibnamefont
  {Diribarne}}, \bibinfo {author} {\bibfnamefont {B.}~\bibnamefont {Rousset}},
  \bibinfo {author} {\bibfnamefont {Y.~A.}\ \bibnamefont {Sergeev}}, \bibinfo
  {author} {\bibfnamefont {J.}~\bibnamefont {Valentin}},\ and\ \bibinfo
  {author} {\bibfnamefont {P.-E.}\ \bibnamefont {Roche}},\ }\bibfield  {title}
  {\bibinfo {title} {Cooling with a subsonic flow of quantum fluid},\ }\href
  {https://doi.org/10.1103/PhysRevB.103.144509} {\bibfield  {journal} {\bibinfo
   {journal} {Phys. Rev. B}\ }\textbf {\bibinfo {volume} {103}},\ \bibinfo
  {pages} {144509} (\bibinfo {year} {2021})}\BibitemShut {NoStop}%
\bibitem [{\citenamefont {Maurer}\ and\ \citenamefont
  {Tabeling}(1998)}]{Maurer98}%
  \BibitemOpen
  \bibfield  {author} {\bibinfo {author} {\bibfnamefont {J.}~\bibnamefont
  {Maurer}}\ and\ \bibinfo {author} {\bibfnamefont {P.}~\bibnamefont
  {Tabeling}},\ }\bibfield  {title} {\bibinfo {title} {Local investigation of
  superfluid turbulence},\ }\href {https://doi.org/10.1209/epl/i1998-00314-9}
  {\bibfield  {journal} {\bibinfo  {journal} {Europhysics Letters ({EPL})}\
  }\textbf {\bibinfo {volume} {43}},\ \bibinfo {pages} {29} (\bibinfo {year}
  {1998})}\BibitemShut {NoStop}%
\bibitem [{\citenamefont {Salort}\ \emph {et~al.}(2010)\citenamefont {Salort},
  \citenamefont {Baudet}, \citenamefont {Castaing}, \citenamefont {Chabaud},
  \citenamefont {Daviaud}, \citenamefont {Didelot}, \citenamefont {Diribarne},
  \citenamefont {Dubrulle}, \citenamefont {Gagne}, \citenamefont {Gauthier},
  \citenamefont {Girard}, \citenamefont {Hébral}, \citenamefont {Rousset},
  \citenamefont {Thibault},\ and\ \citenamefont {Roche}}]{Salort10}%
  \BibitemOpen
  \bibfield  {author} {\bibinfo {author} {\bibfnamefont {J.}~\bibnamefont
  {Salort}}, \bibinfo {author} {\bibfnamefont {C.}~\bibnamefont {Baudet}},
  \bibinfo {author} {\bibfnamefont {B.}~\bibnamefont {Castaing}}, \bibinfo
  {author} {\bibfnamefont {B.}~\bibnamefont {Chabaud}}, \bibinfo {author}
  {\bibfnamefont {F.}~\bibnamefont {Daviaud}}, \bibinfo {author} {\bibfnamefont
  {T.}~\bibnamefont {Didelot}}, \bibinfo {author} {\bibfnamefont
  {P.}~\bibnamefont {Diribarne}}, \bibinfo {author} {\bibfnamefont
  {B.}~\bibnamefont {Dubrulle}}, \bibinfo {author} {\bibfnamefont
  {Y.}~\bibnamefont {Gagne}}, \bibinfo {author} {\bibfnamefont
  {F.}~\bibnamefont {Gauthier}}, \bibinfo {author} {\bibfnamefont
  {A.}~\bibnamefont {Girard}}, \bibinfo {author} {\bibfnamefont
  {B.}~\bibnamefont {Hébral}}, \bibinfo {author} {\bibfnamefont
  {B.}~\bibnamefont {Rousset}}, \bibinfo {author} {\bibfnamefont
  {P.}~\bibnamefont {Thibault}},\ and\ \bibinfo {author} {\bibfnamefont
  {P.-E.}\ \bibnamefont {Roche}},\ }\bibfield  {title} {\bibinfo {title}
  {Turbulent velocity spectra in superfluid flows},\ }\href
  {https://doi.org/10.1063/1.3504375} {\bibfield  {journal} {\bibinfo
  {journal} {Physics of Fluids}\ }\textbf {\bibinfo {volume} {22}},\ \bibinfo
  {pages} {125102} (\bibinfo {year} {2010})}\BibitemShut {NoStop}%
\bibitem [{\citenamefont {Rousset}\ \emph {et~al.}(2014)\citenamefont
  {Rousset}, \citenamefont {Bonnay}, \citenamefont {Diribarne}, \citenamefont
  {Girard}, \citenamefont {Poncet}, \citenamefont {Herbert}, \citenamefont
  {Salort}, \citenamefont {Baudet}, \citenamefont {Castaing}, \citenamefont
  {Chevillard}, \citenamefont {Daviaud}, \citenamefont {Dubrulle},
  \citenamefont {Gagne}, \citenamefont {Gibert}, \citenamefont {Hébral},
  \citenamefont {Lehner}, \citenamefont {Roche}, \citenamefont {Saint-Michel},\
  and\ \citenamefont {Bon~Mardion}}]{Rousset14}%
  \BibitemOpen
  \bibfield  {author} {\bibinfo {author} {\bibfnamefont {B.}~\bibnamefont
  {Rousset}}, \bibinfo {author} {\bibfnamefont {P.}~\bibnamefont {Bonnay}},
  \bibinfo {author} {\bibfnamefont {P.}~\bibnamefont {Diribarne}}, \bibinfo
  {author} {\bibfnamefont {A.}~\bibnamefont {Girard}}, \bibinfo {author}
  {\bibfnamefont {J.~M.}\ \bibnamefont {Poncet}}, \bibinfo {author}
  {\bibfnamefont {E.}~\bibnamefont {Herbert}}, \bibinfo {author} {\bibfnamefont
  {J.}~\bibnamefont {Salort}}, \bibinfo {author} {\bibfnamefont
  {C.}~\bibnamefont {Baudet}}, \bibinfo {author} {\bibfnamefont
  {B.}~\bibnamefont {Castaing}}, \bibinfo {author} {\bibfnamefont
  {L.}~\bibnamefont {Chevillard}}, \bibinfo {author} {\bibfnamefont
  {F.}~\bibnamefont {Daviaud}}, \bibinfo {author} {\bibfnamefont
  {B.}~\bibnamefont {Dubrulle}}, \bibinfo {author} {\bibfnamefont
  {Y.}~\bibnamefont {Gagne}}, \bibinfo {author} {\bibfnamefont
  {M.}~\bibnamefont {Gibert}}, \bibinfo {author} {\bibfnamefont
  {B.}~\bibnamefont {Hébral}}, \bibinfo {author} {\bibfnamefont
  {T.}~\bibnamefont {Lehner}}, \bibinfo {author} {\bibfnamefont {P.-E.}\
  \bibnamefont {Roche}}, \bibinfo {author} {\bibfnamefont {B.}~\bibnamefont
  {Saint-Michel}},\ and\ \bibinfo {author} {\bibfnamefont {M.}~\bibnamefont
  {Bon~Mardion}},\ }\bibfield  {title} {\bibinfo {title} {Superfluid high
  reynolds von kármán experiment},\ }\href
  {https://doi.org/10.1063/1.4897542} {\bibfield  {journal} {\bibinfo
  {journal} {Reviews of Scientific Instruments}\ }\textbf {\bibinfo {volume}
  {85}},\ \bibinfo {pages} {103908} (\bibinfo {year} {2014})}\BibitemShut
  {NoStop}%
\bibitem [{\citenamefont {Salort}\ \emph {et~al.}(2021)\citenamefont {Salort},
  \citenamefont {Chill{\`{a}}}, \citenamefont {Rusaou\"en}, \citenamefont
  {Roche}, \citenamefont {Gibert}, \citenamefont {Moukharski}, \citenamefont
  {Braslau}, \citenamefont {Daviaud}, \citenamefont {Gallet}, \citenamefont
  {Saw}, \citenamefont {Dubrulle}, \citenamefont {Diribarne}, \citenamefont
  {Rousset}, \citenamefont {Bon-Mardion}, \citenamefont {Moro}, \citenamefont
  {Girard}, \citenamefont {Baudet}, \citenamefont {L'vov}, \citenamefont
  {Golov},\ and\ \citenamefont {Nazarenko}}]{Salort21}%
  \BibitemOpen
  \bibfield  {author} {\bibinfo {author} {\bibfnamefont {J.}~\bibnamefont
  {Salort}}, \bibinfo {author} {\bibfnamefont {F.}~\bibnamefont
  {Chill{\`{a}}}}, \bibinfo {author} {\bibfnamefont {E.}~\bibnamefont
  {Rusaou\"en}}, \bibinfo {author} {\bibfnamefont {P.-E.}\ \bibnamefont
  {Roche}}, \bibinfo {author} {\bibfnamefont {M.}~\bibnamefont {Gibert}},
  \bibinfo {author} {\bibfnamefont {I.}~\bibnamefont {Moukharski}}, \bibinfo
  {author} {\bibfnamefont {A.}~\bibnamefont {Braslau}}, \bibinfo {author}
  {\bibfnamefont {F.}~\bibnamefont {Daviaud}}, \bibinfo {author} {\bibfnamefont
  {B.}~\bibnamefont {Gallet}}, \bibinfo {author} {\bibfnamefont {E.-W.}\
  \bibnamefont {Saw}}, \bibinfo {author} {\bibfnamefont {B.}~\bibnamefont
  {Dubrulle}}, \bibinfo {author} {\bibfnamefont {P.}~\bibnamefont {Diribarne}},
  \bibinfo {author} {\bibfnamefont {B.}~\bibnamefont {Rousset}}, \bibinfo
  {author} {\bibfnamefont {M.}~\bibnamefont {Bon-Mardion}}, \bibinfo {author}
  {\bibfnamefont {J.-P.}\ \bibnamefont {Moro}}, \bibinfo {author}
  {\bibfnamefont {A.}~\bibnamefont {Girard}}, \bibinfo {author} {\bibfnamefont
  {C.}~\bibnamefont {Baudet}}, \bibinfo {author} {\bibfnamefont
  {V.}~\bibnamefont {L'vov}}, \bibinfo {author} {\bibfnamefont
  {A.}~\bibnamefont {Golov}},\ and\ \bibinfo {author} {\bibfnamefont
  {S.}~\bibnamefont {Nazarenko}},\ }\bibfield  {title} {\bibinfo {title}
  {Experimental signature of quantum turbulence in velocity spectra?},\ }\href
  {https://doi.org/10.1088/1367-2630/abfe1f} {\bibfield  {journal} {\bibinfo
  {journal} {New Journal of Physics}\ }\textbf {\bibinfo {volume} {23}},\
  \bibinfo {pages} {063005} (\bibinfo {year} {2021})}\BibitemShut {NoStop}%
\bibitem [{\citenamefont {Vallikivi}\ and\ \citenamefont
  {Smits}(2014)}]{Vallikivi14}%
  \BibitemOpen
  \bibfield  {author} {\bibinfo {author} {\bibfnamefont {M.}~\bibnamefont
  {Vallikivi}}\ and\ \bibinfo {author} {\bibfnamefont {A.~J.}\ \bibnamefont
  {Smits}},\ }\bibfield  {title} {\bibinfo {title} {Fabrication and
  characterization of a novel nanoscale thermal anemometry probe},\ }\href
  {https://doi.org/10.1109/JMEMS.2014.2299276} {\bibfield  {journal} {\bibinfo
  {journal} {Journal of Microelectromechanical Systems}\ }\textbf {\bibinfo
  {volume} {23}},\ \bibinfo {pages} {899} (\bibinfo {year} {2014})}\BibitemShut
  {NoStop}%
\bibitem [{\citenamefont {Fan}\ \emph {et~al.}(2015)\citenamefont {Fan},
  \citenamefont {Arwatz}, \citenamefont {Van~Buren}, \citenamefont {Hoffman},\
  and\ \citenamefont {Hultmark}}]{Fan15}%
  \BibitemOpen
  \bibfield  {author} {\bibinfo {author} {\bibfnamefont {Y.}~\bibnamefont
  {Fan}}, \bibinfo {author} {\bibfnamefont {G.}~\bibnamefont {Arwatz}},
  \bibinfo {author} {\bibfnamefont {T.}~\bibnamefont {Van~Buren}}, \bibinfo
  {author} {\bibfnamefont {D.}~\bibnamefont {Hoffman}},\ and\ \bibinfo {author}
  {\bibfnamefont {M.}~\bibnamefont {Hultmark}},\ }\bibfield  {title} {\bibinfo
  {title} {Nanoscale sensing devices for turbulence measurements},\ }\href
  {https://doi.org/10.1007/s00348-015-2000-0} {\bibfield  {journal} {\bibinfo
  {journal} {Experiments in Fluids}\ }\textbf {\bibinfo {volume} {56}},\
  \bibinfo {pages} {1} (\bibinfo {year} {2015})}\BibitemShut {NoStop}%
\bibitem [{\citenamefont {Diribarne}\ \emph {et~al.}(2019)\citenamefont
  {Diribarne}, \citenamefont {Thibault},\ and\ \citenamefont
  {Roche}}]{Diribarne19}%
  \BibitemOpen
  \bibfield  {author} {\bibinfo {author} {\bibfnamefont {P.}~\bibnamefont
  {Diribarne}}, \bibinfo {author} {\bibfnamefont {P.}~\bibnamefont
  {Thibault}},\ and\ \bibinfo {author} {\bibfnamefont {P.-E.}\ \bibnamefont
  {Roche}},\ }\bibfield  {title} {\bibinfo {title} {Nano-shaped hot-wire for
  ultra-high resolution anemometry in cryogenic helium},\ }\href
  {https://doi.org/10.1063/1.5116852} {\bibfield  {journal} {\bibinfo
  {journal} {Review of Scientific Instruments}\ }\textbf {\bibinfo {volume}
  {90}},\ \bibinfo {pages} {105004} (\bibinfo {year} {2019})}\BibitemShut
  {NoStop}%
\bibitem [{\citenamefont {Le-The}\ \emph {et~al.}(2021)\citenamefont {Le-The},
  \citenamefont {K{\"u}chler}, \citenamefont {van~den Berg}, \citenamefont
  {Bodenschatz}, \citenamefont {Lohse},\ and\ \citenamefont {Krug}}]{LeThe21}%
  \BibitemOpen
  \bibfield  {author} {\bibinfo {author} {\bibfnamefont {H.}~\bibnamefont
  {Le-The}}, \bibinfo {author} {\bibfnamefont {C.}~\bibnamefont {K{\"u}chler}},
  \bibinfo {author} {\bibfnamefont {A.}~\bibnamefont {van~den Berg}}, \bibinfo
  {author} {\bibfnamefont {E.}~\bibnamefont {Bodenschatz}}, \bibinfo {author}
  {\bibfnamefont {D.}~\bibnamefont {Lohse}},\ and\ \bibinfo {author}
  {\bibfnamefont {D.}~\bibnamefont {Krug}},\ }\bibfield  {title} {\bibinfo
  {title} {Fabrication of freestanding pt nanowires for use as thermal
  anemometry probes in turbulence measurements},\ }\href
  {https://doi.org/10.1038/s41378-021-00255-0} {\bibfield  {journal} {\bibinfo
  {journal} {Microsystems \& Nanoengineering}\ }\textbf {\bibinfo {volume}
  {7}},\ \bibinfo {pages} {1} (\bibinfo {year} {2021})}\BibitemShut {NoStop}%
\bibitem [{\citenamefont {Dur\`i}\ \emph {et~al.}(2015)\citenamefont {Dur\`i},
  \citenamefont {Baudet}, \citenamefont {Moro}, \citenamefont {Roche},\ and\
  \citenamefont {Diribarne}}]{Duri15}%
  \BibitemOpen
  \bibfield  {author} {\bibinfo {author} {\bibfnamefont {D.}~\bibnamefont
  {Dur\`i}}, \bibinfo {author} {\bibfnamefont {C.}~\bibnamefont {Baudet}},
  \bibinfo {author} {\bibfnamefont {J.-P.}\ \bibnamefont {Moro}}, \bibinfo
  {author} {\bibfnamefont {P.-E.}\ \bibnamefont {Roche}},\ and\ \bibinfo
  {author} {\bibfnamefont {P.}~\bibnamefont {Diribarne}},\ }\bibfield  {title}
  {\bibinfo {title} {Hot-wire anemometry for superfluid turbulent coflows},\
  }\href {https://doi.org/10.1063/1.4913530} {\bibfield  {journal} {\bibinfo
  {journal} {Reviews of Scientific Instruments}\ }\textbf {\bibinfo {volume}
  {86}},\  (\bibinfo {year} {2015})}\BibitemShut {NoStop}%
\bibitem [{\citenamefont {Berberig}\ \emph {et~al.}(1998)\citenamefont
  {Berberig}, \citenamefont {Nottmeyer}, \citenamefont {Mizuno}, \citenamefont
  {Kanai},\ and\ \citenamefont {Kobayashi}}]{Berberig98}%
  \BibitemOpen
  \bibfield  {author} {\bibinfo {author} {\bibfnamefont {O.}~\bibnamefont
  {Berberig}}, \bibinfo {author} {\bibfnamefont {K.}~\bibnamefont {Nottmeyer}},
  \bibinfo {author} {\bibfnamefont {J.}~\bibnamefont {Mizuno}}, \bibinfo
  {author} {\bibfnamefont {Y.}~\bibnamefont {Kanai}},\ and\ \bibinfo {author}
  {\bibfnamefont {T.}~\bibnamefont {Kobayashi}},\ }\bibfield  {title} {\bibinfo
  {title} {The prandtl micro flow sensor (pmfs): a novel silicon diaphragm
  capacitive sensor for flow-velocity measurement},\ }\href
  {https://doi.org/10.1016/S0924-4247(97)01733-0} {\bibfield  {journal}
  {\bibinfo  {journal} {Sensors and Actuators A: Physical}\ }\textbf {\bibinfo
  {volume} {66}},\ \bibinfo {pages} {93} (\bibinfo {year} {1998})}\BibitemShut
  {NoStop}%
\bibitem [{\citenamefont {Moukharski}\ and\ \citenamefont
  {Braslau}(2018)}]{Moukharski18}%
  \BibitemOpen
  \bibfield  {author} {\bibinfo {author} {\bibfnamefont {I.}~\bibnamefont
  {Moukharski}}\ and\ \bibinfo {author} {\bibfnamefont {A.}~\bibnamefont
  {Braslau}},\ }\href {https://patents.google.com/patent/US10018489B2/en}
  {\bibinfo {title} {Miniature differential pressure flow sensor}} (\bibinfo
  {year} {2018}),\ \bibinfo {note} {uS Patent 10,018,489}\BibitemShut {NoStop}%
\bibitem [{\citenamefont {Moukharski}(2019)}]{Moukharski17}%
  \BibitemOpen
  \bibfield  {author} {\bibinfo {author} {\bibfnamefont {I.}~\bibnamefont
  {Moukharski}},\ }\href {https://patents.google.com/patent/US20190138572A1/en}
  {\bibinfo {title} {Method and device for reducing noise in a modulated
  signal}} (\bibinfo {year} {2019}),\ \bibinfo {note} {uS Patent
  20,190,138,572}\BibitemShut {NoStop}%
\bibitem [{\citenamefont {Ravelet}(2005)}]{Ravelet05}%
  \BibitemOpen
  \bibfield  {author} {\bibinfo {author} {\bibfnamefont {F.}~\bibnamefont
  {Ravelet}},\ }\emph {\bibinfo {title} {Bifurcations globales hydrodynamiques
  et magnétohydrodynamiques dans un écoulement de von K\'arm\'an
  turbulent}},\ \href {https://pastel.archives-ouvertes.fr/tel-00011016}
  {\bibinfo {type} {Theses}},\ \bibinfo  {school} {Ecole Polytechnique X}
  (\bibinfo {year} {2005})\BibitemShut {NoStop}%
\bibitem [{\citenamefont {Monchaux}\ \emph {et~al.}(2006)\citenamefont
  {Monchaux}, \citenamefont {Ravelet}, \citenamefont {Dubrulle}, \citenamefont
  {Chiffaudel},\ and\ \citenamefont {Daviaud}}]{Monchaux07}%
  \BibitemOpen
  \bibfield  {author} {\bibinfo {author} {\bibfnamefont {R.}~\bibnamefont
  {Monchaux}}, \bibinfo {author} {\bibfnamefont {F.}~\bibnamefont {Ravelet}},
  \bibinfo {author} {\bibfnamefont {B.}~\bibnamefont {Dubrulle}}, \bibinfo
  {author} {\bibfnamefont {A.}~\bibnamefont {Chiffaudel}},\ and\ \bibinfo
  {author} {\bibfnamefont {F.}~\bibnamefont {Daviaud}},\ }\bibfield  {title}
  {\bibinfo {title} {Properties of steady states in turbulent axisymmetric
  flows},\ }\href {https://doi.org/10.1103/PhysRevLett.96.124502} {\bibfield
  {journal} {\bibinfo  {journal} {Phys. Rev. Lett.}\ }\textbf {\bibinfo
  {volume} {96}},\ \bibinfo {pages} {124502} (\bibinfo {year}
  {2006})}\BibitemShut {NoStop}%
\bibitem [{\citenamefont {Cortet}\ \emph {et~al.}(2010)\citenamefont {Cortet},
  \citenamefont {Chiffaudel}, \citenamefont {Daviaud},\ and\ \citenamefont
  {Dubrulle}}]{Cortet10}%
  \BibitemOpen
  \bibfield  {author} {\bibinfo {author} {\bibfnamefont {P.-P.}\ \bibnamefont
  {Cortet}}, \bibinfo {author} {\bibfnamefont {A.}~\bibnamefont {Chiffaudel}},
  \bibinfo {author} {\bibfnamefont {F.}~\bibnamefont {Daviaud}},\ and\ \bibinfo
  {author} {\bibfnamefont {B.}~\bibnamefont {Dubrulle}},\ }\bibfield  {title}
  {\bibinfo {title} {Experimental evidence of a phase transition in a closed
  turbulent flow},\ }\href {https://doi.org/10.1103/PhysRevLett.105.214501}
  {\bibfield  {journal} {\bibinfo  {journal} {Phys. Rev. Lett.}\ }\textbf
  {\bibinfo {volume} {105}},\ \bibinfo {pages} {214501} (\bibinfo {year}
  {2010})}\BibitemShut {NoStop}%
\bibitem [{\citenamefont {Saint-Michel}\ \emph {et~al.}(2014)\citenamefont
  {Saint-Michel}, \citenamefont {Daviaud},\ and\ \citenamefont
  {Dubrulle}}]{SaintMichel14}%
  \BibitemOpen
  \bibfield  {author} {\bibinfo {author} {\bibfnamefont {B.}~\bibnamefont
  {Saint-Michel}}, \bibinfo {author} {\bibfnamefont {F.}~\bibnamefont
  {Daviaud}},\ and\ \bibinfo {author} {\bibfnamefont {B.}~\bibnamefont
  {Dubrulle}},\ }\bibfield  {title} {\bibinfo {title} {A zero-mode mechanism
  for spontaneous symmetry breaking in a turbulent von k{\'{a}}rm{\'{a}}n
  flow},\ }\href {https://doi.org/10.1088/1367-2630/16/1/013055} {\bibfield
  {journal} {\bibinfo  {journal} {New Journal of Physics}\ }\textbf {\bibinfo
  {volume} {16}},\ \bibinfo {pages} {013055} (\bibinfo {year}
  {2014})}\BibitemShut {NoStop}%
\bibitem [{\citenamefont {Thalabard}\ \emph {et~al.}(2015)\citenamefont
  {Thalabard}, \citenamefont {Saint-Michel}, \citenamefont {Herbert},
  \citenamefont {Daviaud},\ and\ \citenamefont {Dubrulle}}]{Thalabard_2015}%
  \BibitemOpen
  \bibfield  {author} {\bibinfo {author} {\bibfnamefont {S.}~\bibnamefont
  {Thalabard}}, \bibinfo {author} {\bibfnamefont {B.}~\bibnamefont
  {Saint-Michel}}, \bibinfo {author} {\bibfnamefont {E.}~\bibnamefont
  {Herbert}}, \bibinfo {author} {\bibfnamefont {F.}~\bibnamefont {Daviaud}},\
  and\ \bibinfo {author} {\bibfnamefont {B.}~\bibnamefont {Dubrulle}},\
  }\bibfield  {title} {\bibinfo {title} {A statistical mechanics framework for
  the large-scale structure of turbulent von k{\'{a}}rm{\'{a}}n flows},\ }\href
  {https://doi.org/10.1088/1367-2630/17/6/063006} {\bibfield  {journal}
  {\bibinfo  {journal} {New Journal of Physics}\ }\textbf {\bibinfo {volume}
  {17}},\ \bibinfo {pages} {063006} (\bibinfo {year} {2015})}\BibitemShut
  {NoStop}%
\bibitem [{\citenamefont {Salort}\ \emph
  {et~al.}(2011{\natexlab{a}})\citenamefont {Salort}, \citenamefont {Roche},\
  and\ \citenamefont {L{\'e}v{\^e}que}}]{Salort11b}%
  \BibitemOpen
  \bibfield  {author} {\bibinfo {author} {\bibfnamefont {J.}~\bibnamefont
  {Salort}}, \bibinfo {author} {\bibfnamefont {P.-E.}\ \bibnamefont {Roche}},\
  and\ \bibinfo {author} {\bibfnamefont {E.}~\bibnamefont {L{\'e}v{\^e}que}},\
  }\bibfield  {title} {\bibinfo {title} {Mesoscale equipartition of kinetic
  energy in quantum turbulence},\ }\href
  {https://doi.org/10.1209/0295-5075/94/24001} {\bibfield  {journal} {\bibinfo
  {journal} {EPL}\ }\textbf {\bibinfo {volume} {94}},\ \bibinfo {pages} {24001}
  (\bibinfo {year} {2011}{\natexlab{a}})}\BibitemShut {NoStop}%
\bibitem [{\citenamefont {Strouhal}(1878)}]{strouhal1878}%
  \BibitemOpen
  \bibfield  {author} {\bibinfo {author} {\bibfnamefont {V.}~\bibnamefont
  {Strouhal}},\ }\bibfield  {title} {\bibinfo {title} {Ueber eine besondere art
  der tonerregung},\ }\href {https://doi.org/10.1002/andp.18782411005}
  {\bibfield  {journal} {\bibinfo  {journal} {Annalen der Physik und Chemie}\
  }\textbf {\bibinfo {volume} {V}},\ \bibinfo {pages} {216} (\bibinfo {year}
  {1878})}\BibitemShut {NoStop}%
\bibitem [{\citenamefont {Feynman}(1955)}]{Feynman55}%
  \BibitemOpen
  \bibfield  {author} {\bibinfo {author} {\bibfnamefont {R.}~\bibnamefont
  {Feynman}},\ }\bibfield  {title} {\bibinfo {title} {Application of quantum
  mechanics to liquid helium},\ }in\ \href
  {https://doi.org/https://doi.org/10.1016/S0079-6417(08)60077-3} {\emph
  {\bibinfo {booktitle} {Progress in Low Temperature Physics}}},\ \bibinfo
  {series} {Progress in Low Temperature Physics}, Vol.~\bibinfo {volume} {1},\
  \bibinfo {editor} {edited by\ \bibinfo {editor} {\bibfnamefont
  {C.}~\bibnamefont {Gorter}}}\ (\bibinfo  {publisher} {Elsevier},\ \bibinfo
  {year} {1955})\ pp.\ \bibinfo {pages} {17--53}\BibitemShut {NoStop}%
\bibitem [{\citenamefont {Yarmchuk}\ \emph {et~al.}(1979)\citenamefont
  {Yarmchuk}, \citenamefont {Gordon},\ and\ \citenamefont
  {Packard}}]{Yarmchuk79}%
  \BibitemOpen
  \bibfield  {author} {\bibinfo {author} {\bibfnamefont {E.~J.}\ \bibnamefont
  {Yarmchuk}}, \bibinfo {author} {\bibfnamefont {M.~J.~V.}\ \bibnamefont
  {Gordon}},\ and\ \bibinfo {author} {\bibfnamefont {R.~E.}\ \bibnamefont
  {Packard}},\ }\bibfield  {title} {\bibinfo {title} {Observation of stationary
  vortex arrays in rotating superfluid helium},\ }\href
  {https://doi.org/10.1103/PhysRevLett.43.214} {\bibfield  {journal} {\bibinfo
  {journal} {Phys. Rev. Lett.}\ }\textbf {\bibinfo {volume} {43}},\ \bibinfo
  {pages} {214} (\bibinfo {year} {1979})}\BibitemShut {NoStop}%
\bibitem [{\citenamefont {Babuin}\ \emph {et~al.}(2014)\citenamefont {Babuin},
  \citenamefont {Varga}, \citenamefont {Skrbek}, \citenamefont
  {L{\'e}v{\^e}que},\ and\ \citenamefont {Roche}}]{Babuin14}%
  \BibitemOpen
  \bibfield  {author} {\bibinfo {author} {\bibfnamefont {S.}~\bibnamefont
  {Babuin}}, \bibinfo {author} {\bibfnamefont {E.}~\bibnamefont {Varga}},
  \bibinfo {author} {\bibfnamefont {L.}~\bibnamefont {Skrbek}}, \bibinfo
  {author} {\bibfnamefont {E.}~\bibnamefont {L{\'e}v{\^e}que}},\ and\ \bibinfo
  {author} {\bibfnamefont {P.-E.}\ \bibnamefont {Roche}},\ }\bibfield  {title}
  {\bibinfo {title} {Effective viscosity in quantum turbulence: A steady-state
  approach},\ }\href {https://doi.org/10.1209/0295-5075/106/24006} {\bibfield
  {journal} {\bibinfo  {journal} {Europhysics Letters}\ }\textbf {\bibinfo
  {volume} {106}},\ \bibinfo {pages} {24006} (\bibinfo {year}
  {2014})}\BibitemShut {NoStop}%
\bibitem [{\citenamefont {Smith}\ \emph {et~al.}(1993)\citenamefont {Smith},
  \citenamefont {Donnelly}, \citenamefont {Goldenfeld},\ and\ \citenamefont
  {Vinen}}]{Smith93}%
  \BibitemOpen
  \bibfield  {author} {\bibinfo {author} {\bibfnamefont {M.~R.}\ \bibnamefont
  {Smith}}, \bibinfo {author} {\bibfnamefont {R.~J.}\ \bibnamefont {Donnelly}},
  \bibinfo {author} {\bibfnamefont {N.}~\bibnamefont {Goldenfeld}},\ and\
  \bibinfo {author} {\bibfnamefont {W.~F.}\ \bibnamefont {Vinen}},\ }\bibfield
  {title} {\bibinfo {title} {Decay of vorticity in homogeneous turbulence},\
  }\href {https://doi.org/10.1103/PhysRevLett.71.2583} {\bibfield  {journal}
  {\bibinfo  {journal} {Phys. Rev. Lett.}\ }\textbf {\bibinfo {volume} {71}},\
  \bibinfo {pages} {2583} (\bibinfo {year} {1993})}\BibitemShut {NoStop}%
\bibitem [{\citenamefont {Stalp}\ \emph {et~al.}(1999)\citenamefont {Stalp},
  \citenamefont {Skrbek},\ and\ \citenamefont {Donnelly}}]{Stalp99}%
  \BibitemOpen
  \bibfield  {author} {\bibinfo {author} {\bibfnamefont {S.~R.}\ \bibnamefont
  {Stalp}}, \bibinfo {author} {\bibfnamefont {L.}~\bibnamefont {Skrbek}},\ and\
  \bibinfo {author} {\bibfnamefont {R.~J.}\ \bibnamefont {Donnelly}},\
  }\bibfield  {title} {\bibinfo {title} {Decay of grid turbulence in a finite
  channel},\ }\href {https://doi.org/10.1103/PhysRevLett.82.4831} {\bibfield
  {journal} {\bibinfo  {journal} {Phys. Rev. Lett.}\ }\textbf {\bibinfo
  {volume} {82}},\ \bibinfo {pages} {4831} (\bibinfo {year}
  {1999})}\BibitemShut {NoStop}%
\bibitem [{\citenamefont {Barenghi}\ \emph
  {et~al.}(2014{\natexlab{a}})\citenamefont {Barenghi}, \citenamefont
  {Skrbek},\ and\ \citenamefont {Sreenivasan}}]{Barenghi14}%
  \BibitemOpen
  \bibfield  {author} {\bibinfo {author} {\bibfnamefont {C.~F.}\ \bibnamefont
  {Barenghi}}, \bibinfo {author} {\bibfnamefont {L.}~\bibnamefont {Skrbek}},\
  and\ \bibinfo {author} {\bibfnamefont {K.~R.}\ \bibnamefont {Sreenivasan}},\
  }\bibfield  {title} {\bibinfo {title} {Introduction to quantum turbulence},\
  }\href {https://doi.org/10.1073/pnas.1400033111} {\bibfield  {journal}
  {\bibinfo  {journal} {Proceedings of the National Academy of Sciences}\
  }\textbf {\bibinfo {volume} {111}},\ \bibinfo {pages} {4647} (\bibinfo {year}
  {2014}{\natexlab{a}})}\BibitemShut {NoStop}%
\bibitem [{\citenamefont {Salort}\ \emph
  {et~al.}(2011{\natexlab{b}})\citenamefont {Salort}, \citenamefont {Chabaud},
  \citenamefont {L{\'{e}}v{\^{e}}que},\ and\ \citenamefont {Roche}}]{Salort11}%
  \BibitemOpen
  \bibfield  {author} {\bibinfo {author} {\bibfnamefont {J.}~\bibnamefont
  {Salort}}, \bibinfo {author} {\bibfnamefont {B.}~\bibnamefont {Chabaud}},
  \bibinfo {author} {\bibfnamefont {E.}~\bibnamefont {L{\'{e}}v{\^{e}}que}},\
  and\ \bibinfo {author} {\bibfnamefont {P.~E.}\ \bibnamefont {Roche}},\
  }\bibfield  {title} {\bibinfo {title} {Investigation of intermittency in
  superfluid turbulence},\ }\href
  {https://doi.org/10.1088/1742-6596/318/4/042014} {\bibfield  {journal}
  {\bibinfo  {journal} {Journal of Physics: Conference Series}\ }\textbf
  {\bibinfo {volume} {318}},\ \bibinfo {pages} {042014} (\bibinfo {year}
  {2011}{\natexlab{b}})}\BibitemShut {NoStop}%
\bibitem [{\citenamefont {Roche}\ \emph {et~al.}(2009)\citenamefont {Roche},
  \citenamefont {Barenghi},\ and\ \citenamefont {Leveque}}]{Roche09}%
  \BibitemOpen
  \bibfield  {author} {\bibinfo {author} {\bibfnamefont {P.-E.}\ \bibnamefont
  {Roche}}, \bibinfo {author} {\bibfnamefont {C.~F.}\ \bibnamefont
  {Barenghi}},\ and\ \bibinfo {author} {\bibfnamefont {E.}~\bibnamefont
  {Leveque}},\ }\bibfield  {title} {\bibinfo {title} {Quantum turbulence at
  finite temperature: The two-fluids cascade},\ }\href
  {https://doi.org/10.1209/0295-5075/87/54006} {\bibfield  {journal} {\bibinfo
  {journal} {{EPL} (Europhysics Letters)}\ }\textbf {\bibinfo {volume} {87}},\
  \bibinfo {pages} {54006} (\bibinfo {year} {2009})}\BibitemShut {NoStop}%
\bibitem [{\citenamefont {Barenghi}\ \emph
  {et~al.}(2014{\natexlab{b}})\citenamefont {Barenghi}, \citenamefont
  {L{\textquoteright}vov},\ and\ \citenamefont {Roche}}]{Barenghi14b}%
  \BibitemOpen
  \bibfield  {author} {\bibinfo {author} {\bibfnamefont {C.~F.}\ \bibnamefont
  {Barenghi}}, \bibinfo {author} {\bibfnamefont {V.~S.}\ \bibnamefont
  {L{\textquoteright}vov}},\ and\ \bibinfo {author} {\bibfnamefont {P.-E.}\
  \bibnamefont {Roche}},\ }\bibfield  {title} {\bibinfo {title} {Experimental,
  numerical, and analytical velocity spectra in turbulent quantum fluid},\
  }\href {https://doi.org/10.1073/pnas.1312548111} {\bibfield  {journal}
  {\bibinfo  {journal} {Proceedings of the National Academy of Sciences}\
  }\textbf {\bibinfo {volume} {111}},\ \bibinfo {pages} {4683} (\bibinfo {year}
  {2014}{\natexlab{b}})}\BibitemShut {NoStop}%
\bibitem [{\citenamefont {Eltsov}\ \emph {et~al.}(2014)\citenamefont {Eltsov},
  \citenamefont {H{\"a}nninen},\ and\ \citenamefont {Krusius}}]{Eltsov14}%
  \BibitemOpen
  \bibfield  {author} {\bibinfo {author} {\bibfnamefont {V.}~\bibnamefont
  {Eltsov}}, \bibinfo {author} {\bibfnamefont {R.}~\bibnamefont
  {H{\"a}nninen}},\ and\ \bibinfo {author} {\bibfnamefont {M.}~\bibnamefont
  {Krusius}},\ }\bibfield  {title} {\bibinfo {title} {Quantum turbulence in
  superfluids with wall-clamped normal component},\ }\href
  {https://doi.org/10.1073/pnas.1312539111} {\bibfield  {journal} {\bibinfo
  {journal} {Proceedings of the National Academy of Sciences}\ }\textbf
  {\bibinfo {volume} {111}},\ \bibinfo {pages} {4711} (\bibinfo {year}
  {2014})}\BibitemShut {NoStop}%
\bibitem [{\citenamefont {Siggia}(1981)}]{siggia1981numerical}%
  \BibitemOpen
  \bibfield  {author} {\bibinfo {author} {\bibfnamefont {E.~D.}\ \bibnamefont
  {Siggia}},\ }\bibfield  {title} {\bibinfo {title} {Numerical study of
  small-scale intermittency in three-dimensional turbulence},\ }\href
  {https://doi.org/10.1017/S002211208100181X} {\bibfield  {journal} {\bibinfo
  {journal} {Journal of Fluid Mechanics}\ }\textbf {\bibinfo {volume} {107}},\
  \bibinfo {pages} {375} (\bibinfo {year} {1981})}\BibitemShut {NoStop}%
\bibitem [{\citenamefont {Douady}\ \emph {et~al.}(1991)\citenamefont {Douady},
  \citenamefont {Couder},\ and\ \citenamefont
  {Brachet}}]{DouadyFilament:PRL1991}%
  \BibitemOpen
  \bibfield  {author} {\bibinfo {author} {\bibfnamefont {S.}~\bibnamefont
  {Douady}}, \bibinfo {author} {\bibfnamefont {Y.}~\bibnamefont {Couder}},\
  and\ \bibinfo {author} {\bibfnamefont {M.~E.}\ \bibnamefont {Brachet}},\
  }\bibfield  {title} {\bibinfo {title} {Direct observation of the
  intermittency of intense vorticity filaments in turbulence},\ }\href
  {https://doi.org/10.1103/PhysRevLett.67.983} {\bibfield  {journal} {\bibinfo
  {journal} {Phys. Rev. Lett.}\ }\textbf {\bibinfo {volume} {67}},\ \bibinfo
  {pages} {983} (\bibinfo {year} {1991})}\BibitemShut {NoStop}%
\bibitem [{\citenamefont {Baggaley}\ \emph {et~al.}(2012)\citenamefont
  {Baggaley}, \citenamefont {Barenghi}, \citenamefont {Shukurov},\ and\
  \citenamefont {Sergeev}}]{Baggaley_Coherentvortexstructures_EPL2012}%
  \BibitemOpen
  \bibfield  {author} {\bibinfo {author} {\bibfnamefont {A.~W.}\ \bibnamefont
  {Baggaley}}, \bibinfo {author} {\bibfnamefont {C.~F.}\ \bibnamefont
  {Barenghi}}, \bibinfo {author} {\bibfnamefont {A.}~\bibnamefont {Shukurov}},\
  and\ \bibinfo {author} {\bibfnamefont {Y.~A.}\ \bibnamefont {Sergeev}},\
  }\bibfield  {title} {\bibinfo {title} {Coherent vortex structures in quantum
  turbulence},\ }\href {https://doi.org/10.1209/0295-5075/98/26002} {\bibfield
  {journal} {\bibinfo  {journal} {EPL}\ }\textbf {\bibinfo {volume} {98}},\
  \bibinfo {pages} {26002} (\bibinfo {year} {2012})}\BibitemShut {NoStop}%
\bibitem [{\citenamefont {Rusaouen}\ \emph {et~al.}(2017)\citenamefont
  {Rusaouen}, \citenamefont {Rousset},\ and\ \citenamefont
  {Roche}}]{Rusaouen17}%
  \BibitemOpen
  \bibfield  {author} {\bibinfo {author} {\bibfnamefont {E.}~\bibnamefont
  {Rusaouen}}, \bibinfo {author} {\bibfnamefont {B.}~\bibnamefont {Rousset}},\
  and\ \bibinfo {author} {\bibfnamefont {P.-E.}\ \bibnamefont {Roche}},\
  }\bibfield  {title} {\bibinfo {title} {Detection of vortex coherent
  structures in superfluid turbulence},\ }\href
  {https://doi.org/10.1209/0295-5075/118/14005} {\bibfield  {journal} {\bibinfo
   {journal} {{EPL} (Europhysics Letters)}\ }\textbf {\bibinfo {volume}
  {118}},\ \bibinfo {pages} {14005} (\bibinfo {year} {2017})}\BibitemShut
  {NoStop}%
\bibitem [{\citenamefont {Jim{\'e}nez}\ and\ \citenamefont
  {Wray}(1998)}]{jimenez1998characteristics}%
  \BibitemOpen
  \bibfield  {author} {\bibinfo {author} {\bibfnamefont {J.}~\bibnamefont
  {Jim{\'e}nez}}\ and\ \bibinfo {author} {\bibfnamefont {A.~A.}\ \bibnamefont
  {Wray}},\ }\bibfield  {title} {\bibinfo {title} {On the characteristics of
  vortex filaments in isotropic turbulence},\ }\href
  {https://doi.org/10.1017/S0022112098002341} {\bibfield  {journal} {\bibinfo
  {journal} {Journal of Fluid Mechanics}\ }\textbf {\bibinfo {volume} {373}},\
  \bibinfo {pages} {255} (\bibinfo {year} {1998})}\BibitemShut {NoStop}%
\bibitem [{\citenamefont {Vinen}(1957)}]{Vinen57c}%
  \BibitemOpen
  \bibfield  {author} {\bibinfo {author} {\bibfnamefont {W.~F.}\ \bibnamefont
  {Vinen}},\ }\bibfield  {title} {\bibinfo {title} {Mutual friction in a heat
  current in liquid helium ii iii. theory of the mutual friction},\ }\href
  {https://doi.org/10.1098/rspa.1957.0191} {\bibfield  {journal} {\bibinfo
  {journal} {Proceedings of the Royal Society of London. Series A. Mathematical
  and Physical Sciences}\ }\textbf {\bibinfo {volume} {242}},\ \bibinfo {pages}
  {493} (\bibinfo {year} {1957})}\BibitemShut {NoStop}%
\bibitem [{\citenamefont {Van~Sciver}(2012)}]{Vansciver12}%
  \BibitemOpen
  \bibfield  {author} {\bibinfo {author} {\bibfnamefont {S.~W.}\ \bibnamefont
  {Van~Sciver}},\ }\bibinfo {title} {He ii heat and mass transfer},\ in\ \href
  {https://doi.org/10.1007/978-1-4419-9979-5_7} {\emph {\bibinfo {booktitle}
  {Helium Cryogenics}}}\ (\bibinfo  {publisher} {Springer},\ \bibinfo {address}
  {New York},\ \bibinfo {year} {2012})\ pp.\ \bibinfo {pages}
  {227--315}\BibitemShut {NoStop}%
\bibitem [{\citenamefont {Lucas}(1970)}]{Lucas70}%
  \BibitemOpen
  \bibfield  {author} {\bibinfo {author} {\bibfnamefont {P.}~\bibnamefont
  {Lucas}},\ }\bibfield  {title} {\bibinfo {title} {Mutual friction
  measurements in uniformly rotating liquid helium},\ }\href
  {https://doi.org/10.1088/0022-3719/3/5/030} {\bibfield  {journal} {\bibinfo
  {journal} {Journal of Physics C: Solid State Physics}\ }\textbf {\bibinfo
  {volume} {3}},\ \bibinfo {pages} {1180} (\bibinfo {year} {1970})}\BibitemShut
  {NoStop}%
\bibitem [{\citenamefont {Tough}(1982)}]{Tough82}%
  \BibitemOpen
  \bibfield  {author} {\bibinfo {author} {\bibfnamefont {J.~T.}\ \bibnamefont
  {Tough}},\ }\bibinfo {title} {Superfluid turbulence}\ (\bibinfo  {publisher}
  {North-Holland Publishing Company},\ \bibinfo {address} {Amsterdam},\
  \bibinfo {year} {1982})\ Chap.~\bibinfo {chapter} {3}, pp.\ \bibinfo {pages}
  {133--219}\BibitemShut {NoStop}%
\end{thebibliography}%

\end{document}